\documentstyle[prd,aps,eqsecnum,preprint,tighten,floats]{revtex}
\begin{document}
\draft
\preprint{
          UFIFT-HEP-93-18
         }
\title{
       Renormalization group study \\
       of the standard model and its extensions: \\
       II. The minimal supersymmetric standard model.\\
      }

\author{
        D.~J.~Casta\~no\thanks{{\tt castano@ufhepa.phys.ufl.edu}. Address
         after Sept. 1st: M.I.T., Center for Theoretical Physics,
         Cambridge, Ma. 02139.},
        E.~J.~Piard\thanks{{\tt piard@uful07.phys.ufl.edu}. Address after
         Sept. 1st: University of Virginia, Department of Physics,
         Charlottesville, Va. 22901. {\tt piard@landau3.phys.virginia.edu}},
        and P.~Ramond\thanks{{\tt ramond@ufhepa.phys.ufl.edu}}.
       }

\address{
         Institute for Fundamental Theory, Department of Physics, \\
         University of Florida, Gainesville, Florida 32611 USA \\
        }

\date{\today}

\maketitle

\begin{abstract}

In this paper we summarize the minimal supersymmetric standard model
as well as the renormalization group equations of its parameters. We
proceed to examine the feasability of the model when the breaking of
supersymmetry is parametrized by the soft terms suggested by
supergravity theories. In such models, the electroweak symmetry is
exact at tree level and is broken spontaneously at one loop order. We
make the additional assumption that the GUT-inspired relation
$m_b=m_\tau$ be valid at the scale where the gauge coupling constants
unify, which constrains the value of the top quark mass. For all types
of soft breaking terms expected in supergravity theories, we present
the results of numerical runs which yield electroweak breaking at the
required scale.  These yield not only the allowed ranges for the soft
supersymmetry breaking parameters, but also the value of
the supersymmetric partner' masses.  For example in the strict no-scale
model, in which global supersymmetry breaking arises solely from soft
gaugino masses, we find that $M_t$ can be no heavier than $\sim 127$
GeV.

\end{abstract}


\narrowtext

\section{ introduction }
\label{ intro }

In the last few years, it has become apparent, using the ever
increasing accuracy in the measurement of the strong coupling, that
Supersymmetry (SUSY) affords an elegant means to achieve gauge
coupling unification \cite{amaldi,ekn,langluo} at scales consistent
with Grand Unified Theories
(GUTs) \cite{patisalam,geoglash,georgi,frimink,gurasi}.  Whereas in the
Standard Model (SM) the three gauge couplings unify ``two by two''
forming the ``GUT triangle,'' in the simplest Minimal SUSY Extension
of the SM (MSSM) these gauge couplings spectacularly unify at a point
(within the experimental errors in their values).  Given that the
scale of unification in these models is generally above the lower
bound set by proton decay, these supersymmetric grand unified theories
(SUSY-GUTs) have regained increasing interest.  Constraints from the
Yukawa sectors of such models have also yielded interesting
predictions for various low energy parameters including the top quark
mass \cite{summer91,kln,anan,giveon}.

The analyses mentioned above have employed the Renormalization Group (RG).  In
the first part of this two part series \cite{big}, we reviewed the use of the
RG for the purposes stated above in the framework of the SM.  We discussed the
aspects of data extraction involved in arriving at boundary conditions for the
renormalization group equations (RGEs).  In attempting to make the work as
complete as possible, we included all RG $\beta$-functions to 2-loops making no
approximations in the Yukawa sector.  Plots of the evolution of all the SM
parameters to this order were presented. Furthermore, we discussed threshold
effects in the running fermion masses, an analysis which to our knowledge had
not been adequately treated in the literature.

In the present paper we generalize this analysis to the minimal
supersymmetric extension of the standard model. We include the
two loop renormalization group equations for the parameters of the model.
We only consider the case where supersymmetry is broken by soft terms in
the Lagrangean, and  use the renormalization group equations for these
soft breaking parameters at the one loop level.
One of the purposes of this paper is to determine the range of these
parameters which leads to electroweak breaking with the correct value of the
Z-mass. We discuss in some detail the aspects of the
effective one loop potential which are relevant to this breaking.
The mass formulas for the
sfermions, Higgses, charginos, and neutralinos are also presented for
completeness. Various boundary conditions at the unification scale in
these minimal low energy supergravity
models are discussed.  We then describe our numerical procedure.
The treatment of thresholds and the ``special'' form of the $\beta$-functions
needed is then discussed .  Although similar analyses have appeared in the
literature \cite{rossrob,klpny,an,na}, we feel it important to present
our results in a way that directly compares all allowed  sets of values
for the soft supersymmetry breaking parameters. Our
purpose is to use the results of our analysis to find to which extent the low
energy data constrains the types of supersymmetry breakings.
Supersymmetry presents us with a way to extrapolate low energy data
to the deep ultraviolet, where they may be compared with the
underlying theory, be it GUTs and/or Superstrings.

\section{ minimal supersymmetric standard model }
\label{ mssm }

In the minimal supersymmetric extension of the standard model,
every particle has a supersymmetric partner, their spins differing by a
half \cite{nilles}.
In addition, two Higgs fields with opposite hypercharges are needed.
Since the superpotential cannot consist of fields and their complex
conjugates, phenomenology requires two Higgs fields to give mass to the
charge $+2/3$ and $-1/3$, $-1$ sectors, respectively.  Remarkably, two
Higgs fields are also needed for both chiral and $SU(2)$ global anomaly
cancellation.

For renormalizable theories, the superpotential can have at most
degree three interactions.  The superpotential for the MSSM is (suppressing
the $SU(2)$ and Weyl metrics)
\begin{equation}
   W = {\hat{\overline u}} {\bf Y}_u {\hat \Phi}_u {\hat Q} +
       {\hat{\overline d}} {\bf Y}_d {\hat \Phi}_d {\hat Q} +
       {\hat{\overline e}} {\bf Y}_e {\hat \Phi}_d {\hat L} \ ,
\label{superpotential}
\end{equation}
where the hat indicates a chiral superfield and the overline denotes a
left handed $CP$ conjugate of a right handed field,
${\overline\psi}=i\sigma_2\psi_R^*$.  The usual Yukawa interactions are
accompanied by new Yukawa interactions among the scalar quarks and
leptons and the Higgsinos in the supersymmetric Lagrangean.  There are
also new gauge Yukawa interactions involving the gauginos.  The new purely
scalar interactions form the scalar potential which is positive definite
in supersymmetric theories.  The scalar potential will be discussed in a
subsequent section.
A remarkable aspect of supersymmetry is that all these new interactions
require no new couplings.
Table I displays the $SU(3)\times SU(2)\times U(1)$
quantum numbers of the chiral (all left handed) and vector superfields of the
MSSM.

This superpotential exhibits two anomalous chiral global symmetries. The
first, called R-symmetry, is characteristic of supersymmetric theories
with cubic superpotentials;  it implies massless gauginos, it must
be broken; this is simply done by adding gaugino masses, breaking it
down to a discrete R-parity. The other is an exact Peccei-Quinn (PQ)
symmetry which, with this superpotential, can only be  spontaneously
broken at the electroweak scale. This is well known to lead to an axion,
heavy enough to be ruled out by experiment. Hence the superpotential
has to be improved so as to avoid this difficulty. This can be done in
several ways.

\noindent
$\bullet$ The standard way is to break the PQ symmetry explicitly through the
addition of the following supersymmetric-invariant term to the
superpotential
\begin{equation}
   \mu {\hat \Phi}_u {\hat \Phi}_d \ .
\label{pq1}
\end{equation}
This represents the simplest way to avoid the phenomenological
disaster just described.  It is intriguing that, in order to achieve
the required electroweak breaking scale, the value of $\mu$ turns out
to be (as we shall see in the next section)of the same order as the
soft breaking parameters.  At this level, it is not possible to
explain this coincidence, since these terms break different
symmetries.

\noindent
$\bullet$ Alternatively, the $\mu$ can be interpreted dynamically
as the vacuum expectation value of a singlet chiral superfield, $\hat N$,
through the following interactions which also preserve supersymmetry
and break PQ explicitly in the Lagrangean
\begin{equation}
   \lambda {\hat N} {\hat \Phi}_u {\hat \Phi}_d -
   \lambda_0 {\hat N} {\hat N} {\hat N} \ .
\label{singletint}
\end{equation}
The effective $\mu$ can be identified with $\lambda<N>$.
This approach provides a natural explanation for $\mu \sim {\cal O}(M_W)$,
if $\lambda$, $\lambda_0 \sim {\cal O}(g_2)$ \cite{ghkd}.

\noindent
$\bullet$ A third possibility implements non-electroweak spontaneous PQ
symmetry breaking and
leads to an invisible axion, but it needs a special scale of the order
$\sim 10^{11}$ GeV.  This introduces another hierarchy of scales.  This model
is parametrized by the following addition to the superpotential
\begin{equation}
   \lambda {\hat X} {\hat \Phi}_u {\hat \Phi}_d +
   \lambda_0 {\hat X} {\hat Y} {\hat Y} \ .
\label{dfs}
\end{equation}

\noindent
$\bullet$ Finally, one can also break both the PQ symmetry and supersymmetry
by adding to the Lagrangean the explicit soft breaking term,
\begin{equation}
   m_3^2\Phi_u\Phi_d \ .
\label{pq4}
\end{equation}
This term can be put in by hand or generated in the low energy supergravity
model for sufficiently general couplings of supergravity to the standard
model \cite{giumas}.

If the superpotential contains the $\mu$ term, then this soft term
will be generated even for the case of minimal coupling.  In such a
case, it is convenient to set $m_3^2=B\mu$, where $B$ is the soft
bilinear coefficient.  In the following we will adopt the simpler
scheme of the $\mu$ term, leaving the other possibilities to future
investigations.

\begin{table}
\caption{Quantum Numbers}
\begin{tabular}{rrrrrrrrrrrrr}
  & ${\hat Q}$ & ${\hat{\overline u}}$ & ${\hat{\overline d}}$ &
${\hat L}$ & ${\hat{\overline e}}$ & ${\hat\Phi_u}$ & ${\hat\Phi_d}$ &
${\hat g}^A$ & ${\hat W}^a$ & ${\hat B}$ \\
\hline
$U(1)$ & $+{1\over6}$ & $-{2\over3}$ & $+{1\over3}$ & $-{1\over2}$ &
$+1$   & $+{1\over2}$ & $-{1\over2}$ & $0$ & $0$ & $0$ \\
$SU(2)$ & ${\bf 2}$ & ${\bf 1}$ & ${\bf 1}$ & ${\bf 2}$ & ${\bf 1}$ &
${\bf 2}$ & ${\bf 2}$ & ${\bf 1}$ & ${\bf 3}$ & ${\bf 1}$ \\
$SU(3)$ & ${\bf 3}$ & ${\overline{\bf 3}}$ & ${\overline{\bf 3}}$ &
${\bf 1}$ & ${\bf 1}$ & ${\bf 1}$ & ${\bf 1}$ & ${\bf 8}$ &
${\bf 1}$ & ${\bf 1}$ \\
\end{tabular}
\end{table}

\section{ Minimal Low Energy Supergravity Model }
\label{ sugra }

Since no super particles have been observed experimentally, supersymmetry,
if truly present in nature, must be broken.  One way to accomplish this
breaking is to add to the Lagrangean soft breaking terms.

The general soft symmetry breaking potential for the MSSM can be written
(including gaugino mass terms)
\widetext
\begin{mathletters}
\begin{eqnarray}
   V_{soft} &=&  m_{\Phi_u}^2 \Phi_u^\dagger \Phi_u^{}
               + m_{\Phi_d}^2 \Phi_d^\dagger \Phi_d^{}
               + B \mu ( \Phi_u \Phi_d + h.c. ) \nonumber \\
   & &\mbox{}  + \sum_i\Bigl(\;
                 m_{{\tilde Q}_i}^2 {\tilde Q}_i^\dagger {\tilde Q}_i
               + m_{{\tilde L}_i}^2 {\tilde L}_i^\dagger {\tilde L}_i
               + m_{{\tilde{\overline u}}_i}^2 {\tilde{\overline u}}_i^\dagger
                           {\tilde{\overline u}}_i
               + m_{{\tilde{\overline d}}_i}^2 {\tilde{\overline d}}_i^\dagger
                           {\tilde{\overline d}}_i
               + m_{{\tilde{\overline e}}_i}^2 {\tilde{\overline e}}_i^\dagger
                           {\tilde{\overline e}}_i \;\Bigr) \nonumber \\
    & &\mbox{} + \sum_{i,j}\Bigl(\;
                 A_u^{ij}Y_u^{ij}{\tilde{\overline u}}_i\Phi_u{\tilde Q}_j
               + A_d^{ij}Y_d^{ij}{\tilde{\overline d}}_i\Phi_d{\tilde Q}_j
               + A_e^{ij}Y_e^{ij}{\tilde{\overline e}}_i\Phi_d{\tilde L}_j
               + h.c.
                 \;\Bigr) \ , \label{vsoft} \\
    V_{gaugino} &=& {1\over2} \sum_{l=1}^3
                M_l \lambda_l \lambda_l + h.c. \ , \label{vgaugino}
\end{eqnarray}
\end{mathletters}
\narrowtext
\noindent
where $V_{gaugino}$ is the Majorana mass terms for the gaugino fields,
$\lambda_l$ (suppressing the group index), corresponding to $U(1)$,
$SU(2)$, and $SU(3)$, respectively.

We see that there can be sixty-three different soft symmetry breaking
parameters.  In addition, these could each be introduced in the
effective Lagrangean at their own scale! Thus the supersymmetry
breaking section runs the risk of being more complicated than the
standard model itself, which only has a mere eighteen parameters!

It is clear that we need a further organizing principle to describe
supersymmetry breaking. One attractive possibility is suggested by the
coupling of the $N=1$ standard model to $N=1$ supergravity (SUGRA).  The
idea is to have supersymmetry broken in a sector that is uncoupled to
the fields of the standard model, except through the universal
interactions with supergravity. Then the net result of this picture is
to produce in the effective Lagrangean of the minimal low energy
supergravity model, a specific pattern of induced soft breaking terms of
the above form, but with far fewer parameters.

Let us summarize the basic facts of supersymmetry breaking.  From the
supersymmetry algebra, one deduces that spontaneous symmetry breaking
occurs if and only if the vacuum energy is not zero.  In global
supersymmetric theories, the scalar potential is a sum of $F$- and
$D$-terms.  Supersymmetry is spontaneously broken if either the vacuum
values of the $F$-term \cite{orai} or $D$-term \cite{fayetil} are
non-zero.  A consequence of the spontaneous supersymmetry breaking is
a massless fermion in analogy with the breaking of an ordinary global
symmetry.  Spontaneous supersymmetry breaking in supergravity occurs
via the super-Higgs mechanism.  The Nambu-Goldstone fermion, or
Goldstino, associated with the breaking of global supersymmetry is
eaten by the gravitino thereby providing it with a mass. We will
assume that the spontaneous breaking of the local N=1 supersymmetry is
communicated to the ``visible'' sector by weak gravitational
interactions from some ``hidden'' sector. This type of spontaneous
breaking of supergravity manifests itself at low energy as explicit
soft breaking terms of supersymmetry.

The SUGRA Lagrangean is characterized by two arbitrary functions of the
fields, a real function $K$ (the K\"ahler potential) that determines the
kinetic terms of the chiral superfields, and an analytic function
$f_{\alpha\beta}$, transforming as
the symmetric product of the adjoint representation of the gauge group, that
determines the kinetic terms of the gauge fields.  In terms of these, the
supergravity scalar potential is
\begin{eqnarray}
   V_{SUGRA} &=& e^{K/M^2}\left[ \left({\partial W \over\partial
                \phi_i} +
   {1\over M^2}{\partial K \over\partial\phi_i} W\right)
    \left({\partial W^* \over\partial \phi_j^*}
   + {1\over M^2}{\partial K \over\partial\phi_j^*} W^*\right)
   {\partial^2 K \over\partial\phi^i\partial\phi^{j*}}
     -  3 {|W|^2 \over M^2} \right] \nonumber \\
   &&\mbox{}+ {1\over2} f_{\alpha\beta}^{-1} D^\alpha D^\beta \ ,
\label{vsugra}
\end{eqnarray}
where $M = M_{Planck}/\sqrt{8\pi}$ and $D^\alpha$ are auxiliary
fields.  In models with minimal kinetic terms for the chiral
superfields (flat K\"ahler potential, $\partial^2 K
/\partial\phi^i\partial\phi^{j*}=\delta_{ij}$), this leads to a common
(gravitino) mass, $m_0$, for all the scalars of the model.  The
presence of non-minimal gauge kinetic terms
($f_{\alpha\beta}\ne\delta_{\alpha\beta}$) implies non-zero masses,
$M_l$, for the gauginos at the GUT scale, $M_X$.  By further assuming
gauge coupling unification, we can take the three gaugino masses to be
equal.  Furthermore, the trilinear soft couplings $A_u^{ij}$,
$A_d^{ij}$, and $A_e^{ij}$ are all equal to a common value $A_0$.
With minimal chiral kinetic terms, the bilinear soft coupling $B_0$ is
related to $A_0$ as $B_0 = A_0 - m_0$.  This scenario has obvious,
desirable features.  First, it is very predictive since it has a few
parameters accounting for thirty-one new masses.  Second, the
universal nature of the squark and slepton masses at $M_X$ helps to
avoid the appearance of unwanted flavor changing neutral current
(FCNC) effects.  In fact, one could argue that the absence of FCNCs
hints at a universal mass for the scalars.

In this paper, we will not concern ourselves with the exact nature of
the soft breaking terms.  We will leave to others the issue of finding
the exact, more fundamental supergravity models which engenders
particular sets of values.  Our philosophy will be to assume the
presence of these terms and to explore their phenomenological
consequences.  We shall study the effects of varying the values the
soft breaking terms over some interesting ranges and for some
exceptional cases. The scales at which these soft breaking parameters
enter the effective low energy Lagrangean are determined by
theoretical prejudices. In view of the unification of gauge couplings
with supersymmetry, it seems natural to input the three gaugino masses
at the GUT scale. We note that the scales at which the sparticle,
Higgs masses and trilinear couplings enter our Lagrangean could in
principle be anywhere between the Planck and GUT scales.  However,
since we are interested in the rough features of the models, we have
chosen, in the name of simplicity, to enter all the soft breaking
parameters at the same GUT scale.

\section{ radiative electroweak breaking }
\label{ rewb }

 The complete scalar tree level potential now appears as
\begin{equation}
   V = V_F + V_D + V_{soft} \ ,
\label{vscalar}
\end{equation}
where $V_F$ contains the potential contributions from the $F$-terms
\begin{eqnarray}
   V_F &=& | {\tilde{\overline u}} {\bf Y}_u^{} {\tilde Q}
         + \mu\Phi_d |^2
         + | {\tilde{\overline d}} {\bf Y}_d^{} {\tilde Q}
         + {\tilde{\overline e}} {\bf Y}_e^{} {\tilde L}
         + \mu\Phi_u |^2 \nonumber \\
   & & \mbox{} + | {\bf Y}_u^{} {\tilde Q} \Phi_u |^2
         + | {\bf Y}_d^{} {\tilde Q} \Phi_d
         + {\bf Y}_e^{} {\tilde L} \Phi_d |^2 \nonumber \\
   & & \mbox{} + | {\tilde{\overline u}} {\bf Y}_u^{} \Phi_u
         + {\tilde{\overline d}} {\bf Y}_d^{} \Phi_d |^2
         + | {\tilde{\overline e}} {\bf Y}_e^{} \Phi_d |^2 \ ,
\label{vf}
\end{eqnarray}
and $V_D$ contains the potential contributions from the $D$-terms
\begin{eqnarray}
   V_D = {g^{\prime 2}\over2}(\ {1\over6}&&{\tilde Q}_i^\dagger{\tilde Q}_i
           - {2\over3}{\tilde{\overline u}}_i^\dagger{\tilde{\overline u}}_i
           + {1\over3}{\tilde{\overline d}}_i^\dagger{\tilde{\overline d}}_i
           - {1\over2}{\tilde L}_i^\dagger{\tilde L}_i
           +{\tilde{\overline e}}_i^\dagger{\tilde{\overline e}}_i
           + {1\over2}\Phi_u^\dagger\Phi_u
           - {1\over2}\Phi_d^\dagger\Phi_d \ )^2 \nonumber \\
           + {g_2^2\over8}&(&\ {\tilde Q}_i^\dagger{\vec\tau}{\tilde Q}_i
           + {\tilde L}_i^\dagger{\vec\tau}{\tilde L}_i
           + \Phi_u^\dagger{\vec\tau}\Phi_u
           + \Phi_d^\dagger{\vec\tau}\Phi_d \ )^2 \nonumber \\
           + {g_3^2\over8}&(&\ {\tilde Q}_i^\dagger{\vec\lambda}{\tilde Q}_i
           - {\tilde{\overline u}}_i^\dagger{\vec\lambda}^*
             {\tilde{\overline u}}_i
           - {\tilde{\overline d}}_i^\dagger{\vec\lambda}^*
             {\tilde{\overline d}} _i
           \ )^2 \ ,
\label{vd}
\end{eqnarray}
where ${\vec\tau}=(\tau_1,\tau_2,\tau_3)$ are the $SU(2)$ Pauli
matrices and ${\vec\lambda}=(\lambda_1,\dots,\lambda_8)$ are the
Gell-Mann $SU(3)$ matrices.
In general, one must impose constraints on the parameters to avoid charge
and color breaking minima in the scalar potential.  Some {\it necessary}
constraints have been formulated, such as
\begin{mathletters}
\label{acontraints}
\begin{eqnarray}
 A_U^2 &<& 3 ( m_{\tilde Q}^2 + m_{\tilde{\overline u}}^2 + m_{\Phi_u}^2 ) \ ,
   \label{auconstraint}\\
 A_D^2 &<& 3 ( m_{\tilde Q}^2 + m_{\tilde{\overline d}}^2 + m_{\Phi_d}^2 ) \ ,
   \label{adconstraint}\\
 A_E^2 &<& 3 ( m_{\tilde L}^2 + m_{\tilde{\overline e}}^2 + m_{\Phi_d}^2 ) \ .
   \label{aeconstraint}
\end{eqnarray}
\end{mathletters}
\noindent
However, these relations are in general neither {\it sufficient} nor indeed
always {\it necessary} \cite{gunion}.  Their derivation involves very specific
assumptions about the spontaneous symmetry breaking.

An appealing feature of the models we are considering is that they can lead
to the breaking of the electroweak symmetry radiatively
\cite{ibaros,ikkt,agpw,ehnt}.
The one loop
effective Higgs potential in these models can be expressed as the sum of the
tree level potential plus a correction coming from the sum of all one loop
diagrams with external lines having zero momenta
\begin{equation}
   V_{1-loop} =  V_{tree}(\Lambda) + \Delta V_1(\Lambda) \ .
\label{v1loop}
\end{equation}
The right hand side is $\Lambda$-independent up to one loop.
The one loop correction is given by
\begin{eqnarray}
   \Delta V_1(\Lambda) &&=\! {1\over64\pi^2} {\rm Str}\{ {\cal M}^4
                  ( \ln{{\cal M}^2\over\Lambda^2} - {3\over2} )\} \nonumber \\
                  &&=\! {1\over64\pi^2}\!\sum_p (-1)^{2s_{\!p}}(2 s_p\! +\! 1 )
                  m_p^4 ( \ln{m_p^2\over\Lambda^2}\! -\! {3\over2} ),
\label{dv1}
\end{eqnarray}
where ${\cal M}^2$ is the field dependent squared mass matrix of the model
and $m_p$ is the eigenvalue mass of the $p^{th}$ particle of spin $s_p$.
The tree level part of the potential is
%
%
\begin{eqnarray}
   V_{tree}(\Lambda) &=& m_1^2(\Lambda) \Phi_d^\dagger (\Lambda)
                                        \Phi_d^{} (\Lambda)
                   + m_2^2(\Lambda) \Phi_u^\dagger (\Lambda)
                                    \Phi_u^{} (\Lambda)
           + m_3^2(\Lambda) ( \Phi_u (\Lambda) \Phi_d (\Lambda)
                   + h.c. ) \nonumber \\
           &+& {g^{\prime 2}(\Lambda)\over8}( \Phi_u^\dagger (\Lambda)
                                                     \Phi_u^{} (\Lambda)
                   - \Phi_d^\dagger (\Lambda)
                     \Phi_d^{} (\Lambda))^2 \nonumber \\
           &+&  {g_2^2(\Lambda)\over8}( \Phi_u^\dagger (\Lambda)
                                    {\vec\tau}\Phi_u^{} (\Lambda)
                   + \Phi_d^\dagger (\Lambda){\vec\tau}
                     \Phi_d^{} (\Lambda) )^2
\label{vtree}
\end{eqnarray}
where
\begin{eqnarray}
   m_1^2(\Lambda) &=& m_{\Phi_d}^2(\Lambda) + \mu^2(\Lambda) \ , \\
   m_2^2(\Lambda) &=& m_{\Phi_u}^2(\Lambda) + \mu^2(\Lambda) \ , \\
   m_3^2(\Lambda) &=& B(\Lambda) \mu(\Lambda) \ .
\label{mvtree}
\end{eqnarray}
The parameters of the potential are taken as running ones, that is, they vary
with scale according to the renormalization group.  The logarithmic term in
the one loop correction is necessary in making $V_{1-loop}(\Lambda)$
independent of $\Lambda$ to this order (up to non-field dependent terms).

Given the low energy scale of electroweak breaking, we must use the
renormalization group to evolve the parameters of the potential to a
convenient scale such as $M_Z$ (where the experimental values of the
gauge couplings are usually cited) thereby making this leading log
approximation valid.  The exact scale is not critical as long as it is
in the electroweak range.  If we define,

\begin{equation}
   {\overline m}_i^2 = m_i^2 + {\partial\Delta V_1\over\partial v_i^2} \ ,
\label{mbarvtree}
\end{equation}
with $v_1=v_d$, $v_2=v_u$ and
\begin{equation}
   {\partial\Delta V_1\over\partial v_i^2} = {1\over 32\pi^2}\sum_p
   (-1)^{2s_p} ( 2s_p + 1 ) m_p^2 ( \ln{m_p^2\over\Lambda^2} -1 )
   {\partial m_p^2\over\partial v_i^2} \ ,
\label{ddv1dv2}
\end{equation}
then minimization of the potential yields the following two conditions
among its parameters
\begin{equation}
   {1\over2}m_Z^2 = { {\overline m}_1^2 - {\overline m}_2^2 \tan^2\beta \over
                    \tan^2\beta - 1 }  ,
\label{min1}
\end{equation}
where $m_Z^2=(g^{\prime 2}+g_2^2) v^2/2$, $v^2=v_u^2+v_d^2$, and
\begin{equation}
   B \mu = {1\over2}( {\overline m}_1^2 + {\overline m}_2^2 )
                   \sin 2\beta \ ,
\label{min2}
\end{equation}
where $\tan\beta = v_u / v_d$.

Although results based on the tree level potential cannot always be
trusted, one can still use it to get some idea under what conditions
electroweak breaking occurs.  The renormalization group evolution of
$m_{\Phi_u}^2$ (see Appendix A) can be such that it turns negative at low
energies, if the top Yukawa coupling is large enough, whereas $m_{\Phi_d}^2$
runs positive.  From Eq.~(\ref{vtree}), the scale at which breaking occurs
$\Lambda_b$ is set by the condition
\begin{equation}
   m_1^2(\Lambda_b) \ m_2^2(\Lambda_b) - m_3^4(\Lambda_b) = 0 \ .
\label{condition1}
\end{equation}
If the free parameters are adjusted properly, then the correct value of the
$Z^0$ mass ($M_Z=91.17$ GeV) can be achieved.

In the tree level analysis, there is another critical scale
$\Lambda_s$ that must be considered.  It is evident from
Eq.~(\ref{vtree}) that the potential becomes unbounded from
below along the equal field (neutral component) direction, if
\begin{equation}
   m_1^2(\Lambda_s) + m_2^2(\Lambda_s) < 2 m_3^2(\Lambda_s) \ .
\label{condition2}
\end{equation}
Since $m_1^2 m_2^2 - m_3^4 \geq 0$ implies $m_1^2 + m_2^2 \geq 2 m_3^2$,
condition (\ref{condition2}) can only occur at scales lower than condition
(\ref{condition1}), so $\Lambda_s<\Lambda_b$.  From this analysis, one
concludes that the tree level vacuum expectation values (VEVs) of the scalar
fields obtained by minimizing the potential are zero above $\Lambda_b$,
and grow to infinity as one approaches $\Lambda_s$
where the potential becomes unbounded from below.  It follows that the
appropriate scale at which to minimize the tree level potential and evaluate
the VEVs is critical.  This scale must be such that the one loop corrections
of the effective potential may safely be neglected.  Only at such a scale
can the tree level results be trusted.  However, there is more than one scale
involved, and therefore, it is difficult if not impossible to find a scale
at which all logarithms may be neglected.  Indeed, the use of tree level
minimization conditions to compute the VEVs at an arbitrary scale
(e.g., $\Lambda=M_Z$) leads to incorrect conclusions about the regions of
parameter space that yield consistent electroweak breaking scenarios
\cite{gamberini}.  When $\Delta V_1$ is included, however, the value of
$\Lambda$ is not critical as long as it is in the neighborhood of $M_Z$.

Reference~\cite{gamberini} gives a prescription for arriving at a scale
($\hat\Lambda$) at which
the tree level and the one loop effective potential results for the VEVs
agree.  Three qualitatively different cases are considered.
In Ref.~\cite{gamberini}, $M_{SUSY}$ parametrizes the superparticle
thresholds, then the cases can be characterized by the orderings:
(a)~$M_{SUSY}<\Lambda_s<\Lambda_b$,
(b)~$\Lambda_s<M_{SUSY}<\Lambda_b$, and
(c)~$\Lambda_s<\Lambda_b<M_{SUSY}$.  In each case, the prescription is to take
${\hat\Lambda}={\rm max}\{ M_{SUSY},\Lambda_s \}$.  Two cases deserve special
mention.  Case (a) cannot be handled using the tree level analysis because
$v_u,v_d\rightarrow\infty$ near $\Lambda_s$.  Fortunately, phenomenological
bounds rule this case out anyway.  In case (c), there is actually no
electroweak breaking.  For scales below $M_{SUSY}$, the superparticles have
decoupled and the effective theory is not supersymmetric.  Therefore, the
running mass parameters of the potential freeze into their values at
$\Lambda=M_{SUSY}$ at which scale there is no electroweak breaking.  Finally,
it must be emphasized that the apparent violent behavior of the VEVs with
scale in the tree level analysis is an artifact of the approximation.
The only physical potential is the full effective potential, and it either
breaks electroweak symmetry or not.  If it does, then the scalar fields have
non-zero VEVs, and these VEVs are non-zero over all scales varying according
to the anomalous dimension of their respective scalar fields (in the Landau
gauge).

In contrast to the tree level potential, the one loop effective potential is
constant against the renormalization group to this order around the
electroweak scale.  The exact scale at which to minimize is no longer
critical.  Moreover the assessment of the masses of the Higgs bosons
based on the one loop effective potential is more accurate.  The tree
level restriction $M_h<M_Z$ is known not to be valid when one loop
corrections, which are large because $M_t$ is large, are included in the
determination of $M_h$ \cite{oyy,haber,erz}.

In this paper, we do not rely on the tree level analysis, rather
we incorporate the one loop corrections.  We include the
dominant contributions from the third family, that is, those of the top
and stop, bottom and sbottom, and tau and stau \cite{ellis,drees}.  We choose
the $Z^0$ mass as the scale at which to evaluate the minimization conditions.
Equations (\ref{min1}) and (\ref{min2}) can be written
\begin{mathletters}
\begin{eqnarray}
   \mu^2(M_Z) &=& {{\overline m}_{\Phi_d}^2 -
   {\overline m}_{\Phi_u}^2 \tan^2\beta \over \tan^2\beta - 1 }
   - {1\over2} m_Z^2 \ , \label{min1a} \\
   B(M_Z) &=& { ({\overline m}_1^2 + {\overline m}_2^2)\sin2\beta \over
              2 \mu(M_Z) } \ , \label{min2a}
\end{eqnarray}
\end{mathletters}
where ${\overline m}_{\Phi_{u,d}}^2 = m_{\Phi_{u,d}}^2 +\partial\Delta
V_1/\partial v_{u,d}^2$ and used to solve for $\mu(M_Z)$ and $B(M_Z)$ given
the value of all the relevant parameters at $M_Z$.  We note that the form of
Eq.~(\ref{min1a}) does not fix the sign of $\mu$, and a choice for its
sign must be made ($\mu$ is multiplicatively renormalized; see Appendix A).
The right hand sides of these equations implicitly involve the VEV at $M_Z$.
In a consistent scenario it would have the value of $v(M_Z)=174.1$ GeV.
If the parameters are such that the right hand side of (\ref{min1a}) is
negative, then the scenario is inconsistent and the electroweak symmetry
fails to be broken.

The validity of using (\ref{v1loop}) at $M_Z$ hinges on the assumption
that there is spontaneous symmetry breaking and that $M_Z=91.17$ GeV.
Given a set of values for the input parameters (soft terms, etc.) at
$M_X$, we proceed by assuming valid electroweak breaking ({\it i.e.},
$M_Z=91.17$ GeV).  The one loop effective potential of
Eq.~(\ref{v1loop}) should then be perturbatively valid at $M_Z$ with
all its running parameters evaluated at this scale.  In this way, we
are renormalization-group-improving the potential.
The validity of this assumption is tested by the consistency of
Eqs.~(\ref{min1a}) and (\ref{min2a}).  Failure to attain consistency
invalidates the initial assumption that the given input values at
$M_X$ can accommodate a low energy world as we know it.

\section{ sparticle masses }

In the following, we list the tree level mass formulas for the
different superpartners.

\subsection{ Sfermion masses }
\label{ sfermion }

The mass matrices for scalar matter are constructed from Eq.(\ref{vscalar}).
For example, in the up squark sector the relevant mass matrix appears
as
\begin{equation}
   {\cal M}_{\tilde u}^2 = \left( \begin{array}{cc}
                               M_{L_i L_j}^2 & M_{L_i R_j}^2 \\
                               M_{R_i L_j}^2 & M_{R_i R_j}^2
                               \end{array} \right)
\label{msup}
\end{equation}
where $i,j = 1,2,3$ are flavor indices and
\begin{mathletters}
\label{mij}
\begin{eqnarray}
   M_{L_i L_j}^2 &=& m_{Q_i}^2 \delta_{ij}
                 + v_u^2 ( {\bf Y}_u^\dagger{\bf Y}_u^{} )_{ij}
                 - {1\over2}( v_d^2 - v_u^2 ) ( Y(u_L)g^{\prime 2}
                 - T_3(u_L) g_2^2 )\delta_{ij} \ , \label{mll} \\
   M_{R_i R_j}^2 &=& m_{u_i}^2 \delta_{ij}
                 + v_u^2 ( {\bf Y}_u^\dagger{\bf Y}_u^{} )_{ij}
                 - {1\over2}( v_d^2 - v_u^2 ) ( Y(u_R)
                   g^{\prime 2} )\delta_{ij} \ , \label{mrr} \\
   M_{R_i L_j}^2 &=& \mu v_d Y_u^{ij} + v_u A_u^{ij} Y_u^{ij} , \
                   \label{mrl} \\
   M_{L_i R_j}^2 &=& M_{R_j L_i}^{2*} \ . \label{mlr}
\end{eqnarray}
\end{mathletters}
\noindent
Note in Table I that $Y=Q-T_3$ in our notation.  Similar matrices follow
for the other sfermions.  These mass formulas as well as the ones to follow
are given in terms of running parameters.  The domain of validity of these
formulas is at low energies ($\sim M_Z$) with the parameters taking on
their renormalization group evolved values at this scale.

\subsection{ Higgs masses }
\label{ higgs }

If we employ the notation
\begin{equation}
   \Phi_1 = \left( \begin{array}{c} \phi_1^0 \\ \phi_1^-
                   \end{array} \right) , \
   \Phi_2 = \left( \begin{array}{c} \phi_2^+ \\ \phi_2^0
                   \end{array} \right) \ ,
\end{equation}
then the  physical masses of the Higgs at tree level are calculated from the
following three matrices
\begin{mathletters}
\begin{eqnarray}
   {1\over2}{\partial^2V_{tree}\over\partial(\Im\phi_i^0)\partial(\Im\phi_j^0)}
   &=& {1\over2} M_A^2 \sin2\beta \left( \begin{array}{cc}
             \tan\beta & 1 \\
             1 & \cot\beta
             \end{array} \right) \label{mhcpodd} \\
   {1\over2}{\partial^2V_{tree}\over\partial(\Re\phi_i^0)\partial(\Re\phi_j^0)}
   &=& {1\over2} M_A^2 \sin2\beta \left( \begin{array}{cc}
             \tan\beta & -1 \\
             -1 & \cot\beta
             \end{array} \right)
      + {1\over2} m_Z^2 \sin2\beta \left( \begin{array}{cc}
             \cot\beta & -1 \\
             -1 & \tan\beta
             \end{array} \right) \label{mhcpeven} \\
   {\partial^2V_{tree}\over\partial(\phi_i^-)\partial(\phi_j^+)}
   &=& {1\over2}M_{H_\pm}^2 \sin2\beta \left( \begin{array}{cc}
             \tan\beta &  1 \\
             1 & \cot\beta
             \end{array} \right) \label{mhcharged}
\label{mh}
\end{eqnarray}
\end{mathletters}
where $M_A^2=m_1^2+m_2^2$, $M_{H_\pm}^2=M_A^2+m_W^2$, and
$m_W^2=g_2^2 v^2 /2$.  The eigenvalues for the first matrix are $0$
corresponding to the Goldstone boson and $M_A^2$ corresponding to the
$CP$ odd scalar.  The second matrix gives the masses of the light and
heavy Higgs bosons
\begin{equation}
M_{H,h}^2={1\over2}[\ (M_A^2 + m_Z^2).
              \pm  \sqrt{(M_A^2+m_Z^2)^2 -
              4M_A^2m_Z^2\cos^22\beta}\ ] \ .
\label{mhiggs}
\end{equation}
The mixing angle $\alpha$ that diagonalizes the matrix (\ref{mhcpeven})
can be expressed
\begin{equation}
   \tan 2\alpha = {(M_A^2+m_Z^2)\over(M_A^2-m_Z^2)} \tan 2\beta \ .
\label{adiag}
\end{equation}
If $M_A^2\gg m_z^2$ which is the limit where the heavy Higgs is very heavy,
this angle coincides with $\beta$.
This tree level result predicts $M_h<M_Z$.  One loop calculations show
that this need not be the case \cite{oyy,haber,erz}.
The third matrix has eigenvalues $0$ and
$M_{H_\pm}^2$ corresponding to a massless, charged Goldstone boson and a
charged scalar.

Including Eq.~(\ref{dv1}) in the calculations leads to corresponding one loop
versions of these masses \cite{ellis,drees}.

\subsection{ Chargino masses }
\label{ chargino }

The following four terms contribute to the chargino masses
\begin{equation}
   -i{\sqrt2}g_2\Phi_u^\dagger{\tau^i\over2}{\tilde\Phi}_u{\tilde W}^i
   -i{\sqrt2}g_2\Phi_d^\dagger{\tau^i\over2}{\tilde\Phi}_d{\tilde W}^i
   - \mu {\tilde\Phi}_u {\tilde\Phi}_d
   + {1\over2}M_2 {\tilde W}^i {\tilde W}^i
   + h.c. \ .
\label{charginomassterms}
\end{equation}
The first two terms are the supersymmetric Yukawa-gauge terms.  Letting
$\lambda^\pm = ({\tilde W}_2 \pm i{\tilde W}_1)/{\sqrt 2}$, the mass matrix
follows
\widetext
\begin{equation}
   \left( \begin{array}{cccc} \lambda^+ & {\tilde\phi}_u^+
                            & \lambda^- & {\tilde\phi}_d^-
          \end{array} \right)
   {1\over2} \left( \begin{array}{cccc} 0 & 0 & M_2 & -g_2v_d \\
                                        0 & 0 & g_2v_u & -\mu \\
                                        M_2 & g_2v_u & 0 & 0 \\
                                        -g_2v_d & -\mu & 0 & 0
                    \end{array} \right)
    \left( \begin{array}{c} \lambda^+ \\ {\tilde\phi}_u^+ \\
                           \lambda^- \\ {\tilde\phi}_d^-
          \end{array} \right).
\label{charginomatrix}
\end{equation}
Diagonalization yields two charged Dirac fermions, ${\tilde C}_1$,
${\tilde C}_2$, with masses
\begin{equation}
   M_{{\tilde C}_{1,2}} = {1\over2}[ ( M_2^2 + \mu^2
   + 2m_W^2) \mp \sqrt{ ( M_2^2 + \mu^2 + 2m_W^2)^2
   - 4 ( M_2 \mu - m_W^2 \sin2\beta)^2 } ] \ .
\label{charginomass}
\end{equation}
\narrowtext

\subsection{ Neutralino masses }
\label{ neutralino }

Contributing to the neutralino masses are the terms in
Eq.~(\ref{charginomassterms}) and
\begin{equation}
   -ig^\prime {\sqrt2} \Phi_u^\dagger (+{1\over2}) {\tilde\Phi_u} {\tilde B}
   -ig^\prime {\sqrt2} \Phi_d^\dagger (-{1\over2}) {\tilde\Phi_d} {\tilde B}
   + {1\over2} M_1 {\tilde B}{\tilde B} \ .
\label{neutralinomassterms}
\end{equation}
The neutralino mass matrix follows
\widetext
\begin{equation}
   \left( \begin{array}{cccc} i{\tilde B} & i{\tilde W}_3 &
                              {\tilde\phi}_d^0 & {\tilde\phi}_u^0
          \end{array} \right) {1\over2}
   \left( \begin{array}{cccc}
          -M_1 & 0 & g^\prime v_d/{\sqrt2} & -g^\prime v_u/{\sqrt2} \\
          0 & -M_2 & -g_2 v_d/{\sqrt2} & g_2 v_u/{\sqrt2} \\
          g^\prime v_d/{\sqrt2} & -g_2 v_d/{\sqrt2} & 0 & \mu \\
          -g^\prime v_u/{\sqrt2} & g_2 v_u/{\sqrt2} & \mu & 0
          \end{array} \right)
   \left( \begin{array}{c} i{\tilde B} \\ i{\tilde W}_3 \\
                           {\tilde\phi}_d^0 \\ {\tilde\phi}_u^0
          \end{array} \right) \ .
\label{neutralinomatrix}
\end{equation}
\narrowtext

Its eigenvalues are the masses of the four neutralinos, ${\tilde
N}_1$, ${\tilde N}_2$, ${\tilde N}_3$, ${\tilde N}_4$.  Typically not
all the eigenvalues of this mass matrix have the same sign.  It was
noted in \cite{marram} that over most of the range of the parameters,
one of the eigenvalues is of the opposite sign compared to the others.
This sign is important in distinguishing this neutralino (the
``flippino'') and especially in deriving mass sum rules \cite{marram}.

\section{ boundary conditions at $M_X$ }
\label{ bc }

In this paper, as in the previous one \cite{big}, we work in the modified
minimal subtraction scheme (${\overline{\rm MS}}$) of renormalization.  The
parameters of the Lagrangean are not in general equal to any corresponding
physical constant.  For example, in the case of masses, except for those of
the bottom and top quark (see \cite{big}), all other physical masses, $M$,
will be determined from their corresponding running masses by the relation
\begin{equation}
   M = m(\Lambda){\vert}_{\Lambda=M} \ .
\label{eq14}
\end{equation}
This equation is easily solved in the course of an integration of the RGEs
for the different masses by noting the scale at which it is valid.
We have collected the renormalization group $\beta$-functions of the MSSM
for the gauge and Yukawa couplings to two loops without making any
approximations in the Yukawa sector in Appendix A.  Also included are the
one loop $\beta$-functions for the soft breaking terms.

Because SUGRA models make simplifying predictions about the soft
parameters at some large scale, we initiate the evolution of the
renormalization group equations at this scale.  For simplicity, we
take this scale to be the gauge unification scale where we expect the
gaugino masses to be equal to some common value.  It has been
demonstrated \cite{amaldi} that the introduction of supersymmetry
leads to gauge coupling unification at approximately $\sim 10^{16}$
GeV.  Therefore we take $M_X=10^{16}$ GeV, and evolve down to $1$ GeV,
the conventional scale at which the running quark masses are given
\cite{gassleut}.  Furthermore, as already discussed, the Higgs
potential must be analyzed at some low energy scale that we choose to
be $M_Z$.

At the unification scale, $M_X$, all the scalars will have a common mass,
$m_0$,
\begin{mathletters}
\begin{eqnarray}
   m_{{\tilde Q}_i}(M_X) &=& m_{{\tilde{\overline u}}_i}(M_X) =
       m_{{\tilde{\overline d}}_i}(M_X) = m_{{\tilde L}_i}(M_X)
   \nonumber \\
           &=& m_{{\tilde{\overline e}}_i}(M_X) = m_{\Phi_u}(M_X) =
       m_{\Phi_d}(M_X)  \nonumber \\
           & \equiv & m_0 \ ,
   \label{bcm0}
\end{eqnarray}
\end{mathletters}
the gauginos will also have a common mass, $m_{1/2}$,
\begin{equation}
   M_1(M_X) = M_2(M_X) = M_3(M_X) \equiv  m_{1/2} \ .
\label{bcmhalf}
\end{equation}
The prefactors of the trilinear soft scalar terms are equal to $A_0$ at $M_X$
\begin{equation}
   A_u^{ij}(M_X) = A_d^{ij}(M_X) = A_e^{ij}(M_X) \equiv A_0 \ .
\label{bca}
\end{equation}
Also we define the bilinear soft scalar coupling and the mixing mass at $M_X$
by $B(M_X) \equiv B_0$, and $\mu(M_X) \equiv \mu_0$.

Furthermore, to constrain the parameter space, we will take the bottom and
tau masses equal at $M_X$
\begin{equation}
   m_b(M_X) = m_\tau(M_X) \ .
\label{mbeqmtau}
\end{equation}
This being the best motivated mass relation in supersymmetric grand
unified theories \cite{massrel}.


The purpose of our present analasys is to determine the allowed range
of the input parameters, $A_0$, $m_0$, $m_{1/2}$, and $\tan\beta$,
which reproduce the known low energy physics. Within this general set
of solutions, there are four subclasses of soft supersymmetry breaking
which are of particular interest.  Three of these have various soft
parameters equal to zero at $M_X$.  The fourth predicts a definite
relationship among three of these parameters.  The first class of
models follows from the no-scale model \cite{noscale} and has
$A_0=m_0=B_0=0$ (strict no-scale model).  In these models, only
gaugino masses provide global supersymmetry breaking.  The second
class is the less constraining no-scale case that has only
$A_0=m_0=0$.  A third class we consider with $A_0=B_0=0$ comes from
some string derived models.  Minimal SUGRA models with canonical
kinetic terms for the chiral superfields form another class and have
$B_0=A_0-m_0$.  Table II lists these four possibilities.  In both
strict no-scale and string inspired cases, we must have $B_0=0$.
However, since for us $B_0$ is an output rather than an input
variable, $B_0=0$ results must be inferred from its behavior upon
varying other input parameters.

\begin{table}
\caption{Models}
\begin{tabular}{lcccc}
                & $A_0$ & $m_0$ & $m_{1/2}$ & $B_0$ \\
\hline
General         & any   & any   & any       & any   \\
Strict No-scale & $0$   & $0$   & any       & $0$   \\
No-scale        & $0$   & $0$   & any       & any   \\
String Inspired & $0$   & any   & any       & $0$   \\
Minimal SUGRA   & any   & any   & any       & $A_0-m_0$   \\
\end{tabular}
\end{table}

\section{ numerical procedure }
\label { numproc }

Our numerical procedure relies on the Runge-Kutta technique of numerical
integration.  To make the process of initialization, or the determination
of all parameters at one scale, more transparent, it is useful to think
of the Runge-Kutta integration of the renormalization group equations
as a vector valued function of a vector variable.  The input vector is the
N-plet of values for all the (N) parameters at $M_X$.  The Runge-Kutta
function then returns an N-plet result representing the running values of
all the parameters at final scales such as $M_Z$ or $1$ GeV.

In order to make a run (using Runge-Kutta) starting from $M_X$, we must have
the values of all the parameters at this scale, but this is difficult to
achieve.  The problem is that
the values of many parameters are known experimentally at low scales.  Also,
relations (\ref{min1a}) and (\ref{min2a}) coming from the one loop effective
potential hold at such low energies.  However, the values of other parameters,
such as soft breaking terms, are most easily understood at higher energies
where theoretical simplification ({\it e.g.}, universality) may be invoked.
We therefore have no scale at which there is both theoretical simplicity and
experimental data.

We can phrase the problem in another way.  Choosing $M_X$
(where we have theoretical simplicity) as our starting point, can we find the
$M_X$ values of all parameters such that we recover the expected low energy
values after renormalization group evolution to experimental scales, and all
constraint relations among the parameters are satisfied at the appropriate
scales?  We distinguish between constrained and free parameters.  The former
are constained by experiment ({\it e.g.}, quark masses, gauge couplings,
etc.) or relations among themselves ({\it e.g.}, bottom $\tau$ Yukawa
equality at $M_X$, minimization conditions at $M_Z$, etc.).
Given the present experimental data, the latter cannot be constrained by
these two criteria and must be viewed as input parameters.
After we have run, we find some ranges of their values to be inconsistent with
phenomenological considerations such as electroweak symmetry breaking and
the resulting sparticle spectrum.
The exact values of the constrained parameters at $M_X$ are affected by
the choice of values for the free parameters at $M_X$ since the evolution
of all parameters are coupled.  Therefore, given a choice for the free
parameters at $M_X$, we must find the $M_X$ values of the constrained
parameters consistent with all the constraints.

The functional nature of the (Runge-Kutta) integration allows us to define a
set of $n$ (where $n$ is the total number of constrained parameters) equations
\begin{equation}
   G_k({\bf x}_0)=0 \ ,
\label{shooting}
\end{equation}
where $k=1,\dots,n$, and ${\bf x}_0$ represents the values of all
constrained parameters at $M_X$.  The solution to the initialization problem,
therefore is reduced to solving a system of $n$ simultaneous non-linear
equations in $n$ unknowns.  In this work as in Ref.~\cite{big}, we will use
routines, based on the ``shooting'' method, to solve systems of nonlinear
equations.  The method involves making a guess for the solution, then
assessing its merits based on how well the equations are satisfied, given some
tolerance.  The next guess (or shot) is adjusted according to how accurate the
previous one was.  The process is optimized and iterated until the routine
converges on a solution.
In this way, the values of all parameters are ascertained at one common
scale which we take to be $M_X$.  When the parameters are evolved to
lower scales using these initial values, they attain the experimentally
known values and they satisfy any relations amongst themselves that
were used as constraints in the shooting procedure.

As discussed previously, we start our runs at $M_X$ at which scale
we can make simplifying assumptions about the soft breaking terms based
on various SUGRA models.  This requires that we use the solution routines
to consistently find the $M_X$ values of all known low energy parameters
such as lepton and quark masses and mixing angles and gauge couplings.
This amounts to solving for sixteen unknowns (nine masses, three angles and
a phase, and three gauge couplings).  Alternatively, we could
start our runs at $M_Z$ or $1$ GeV; however, this now requires solving
for sixty-three unknowns (the values of the soft breaking terms at low
energy) that must evolve to just four different values at $M_X$.  The
efficiency of the former method is obvious.

There are ostensibly seven parameters in the models we consider.  These are
$A_0$, $B_0$, $m_0$, $m_{1/2}$, $\mu_0$, $\tan\beta$, and $m_t$.  The two
minimization constraints (\ref{min1}) and (\ref{min2}) reduce this
set to five, which are taken to be $A_0$, $m_0$, $m_{1/2}$, $\tan\beta$, and
$m_t$.  In the present framework, $B_0$ and $\mu_0$ will be determined
using the numerical solutions routines in
conjunction with the minimization of the one loop effective potential at
$M_Z$ in the process of evolving from $M_X$ to $1$ GeV.  Minimization at $M_Z$
will give $B(M_Z)$ and $\mu(M_Z)$.  To arrive at $B_0$ and $\mu_0$
(their corresponding values at $M_X$), we employ the solution routine as
follows.  A guess for $B_0$ and $\mu_0$ is made at $M_X$ and then the
parameters of the model are run to $M_Z$ at which scale the evolved value of
$B$ is compared to the minimization output value for $B$ at $M_Z$.  The same
is done for $\mu$.  If the compared values agree to some set accuracy,
then $B_0$ and $\mu_0$ are the required values.  Other analyses that also
extract $B(M_Z)$ and $\mu(M_Z)$ simply evolve these two parameters via
their renormalization group equations back to $M_X$ to find $B_0$ and
$\mu_0$ relying on their near decoupling from the full set of RGEs.
We note that the sign of $\mu$ is not determined from the minimization
procedure, thus we must make a choice for it.  To constrain the parameter
space further, the bottom quark and
tau lepton masses will be taken equal at $M_X$.  This equality is a
characteristic of many SUSY-GUTs.  This constrains the model to four free
parameters, $A_0$, $m_0$, $m_{1/2}$, and $\tan\beta$.  Demanding  that
$m_b(M_X)=m_\tau(M_X)$ and achieving the correct physical
masses for the bottom quark and tau lepton fixes the mass of the top quark
which affects the evolution of the bottom Yukawa significantly.
We shall assume gauge coupling unification, an assumption which appears
reasonable when one considers SUSY models with SUSY breaking scales
$\alt 10$ TeV.

In a complete treatment, the solution routines would be used to find
the precise (similar) values of $\alpha_1$, $\alpha_2$, and $\alpha_3$
at $M_X$ that will evolve to the experimentally known values at $M_Z$,
however this increases the CPU time considerably.  We shall therefore
sacrifice some precision in their $M_Z$ values by taking them exactly
equal at $M_X$.  This is already a theoretical oversimplification
since one does not expect the gauge couplings to be exactly equal due
to threshold effects at the GUT scale \cite{fgm}.  We find that for
all cases we have studied, the common value
$\alpha_1^{-1}(M_X)=\alpha_2^{-1}(M_X)=\alpha_3^{-1}(M_X)=25.31$ leads
to errors no bigger than $1\%$, $5\%$, and $10\%$ in $\alpha_1(M_Z)$,
$\alpha_2(M_Z)$, and $\alpha_3(M_Z)$, respectively.  This is not so
bad considering that the (combined experimental and theoretical)
errors on $\alpha_3(M_Z)$ from some processes can be as large as
$10\%$ \cite{big}.

It is well known that there is a fine tuning problem inherent in the
radiatively induced electroweak models.  For certain values of the
parameters, the top quark mass must be tuned to an ``unnaturally'' high
degree of accuracy to achieve the correct value of $M_Z$.  This problem is
generally handled by rejecting models that require ``too much'' tuning.  The
amount of tuning is usually defined quite arbitrarily.  The usual procedure
is to define fine tuning parameters \cite{barbigiudi}
\begin{equation}
   c_i = | {x_i^2\over M_Z^2} {\partial M_Z^2\over\partial x_i^2} | \ ,
\label{ftparam}
\end{equation}
where $x_i$ are parameters of the theory such as $m_0$, $m_{1/2}$,
$\mu$, or $m_t$.  One then demands that the $c_i$ be less than some
chosen value that is typically taken to be $10$.

We have analyzed to
some extent the differences in using the tree level {\it vs.} one loop
effective potential.  The basis for the ``theoretical'' fine tuning problem
can be seen (for example, assuming the renormalization group equation for
the up sector Higgs mass is dominated by the top quark Yukawa coupling)
in the dependence of $M_W$ on the top quark Yukawa coupling $y_t$ \cite{nano}
\begin{equation}
   M_W \sim M_X e^{-1/y_t^2} \ .
\label{finetuning}
\end{equation}
We remark that this fine tuning problem is exacerbated if one uses only the
tree level analysis of the potential.  The vacuum expectation value coming
from the minimization conditions
of the tree level potential changes rapidly from $0$ to infinity over the
interval $(\Lambda_s,\Lambda_b)$.  Using the prescription of
Ref.\cite{gamberini} for the scale ${\hat \Lambda}$ at which to adequately
minimize the tree level potential, to extract $v({\hat\Lambda})$,
and thereby to arrive at a value for $M_Z$, one finds that although a small
variation in $y_t(M_X)$ may lead to a small variation in $\Lambda_b$, the
steepness in the tree level VEV can lead to a large variation in the value of
$v({\hat\Lambda})$ and therefore in $M_Z$.  Hence, in the tree level analysis,
solutions which may be within the bounds of the ``theoretical'' fine tuning
may nevertheless display a fine tuning aspect because of this ``tree level''
fine tuning of $\hat\Lambda$.

However, our use of the one loop effective potential
(i.e., including $\Delta V_1$) stabilizes the VEV around the $M_Z$ scale
and this particular fine tuning goes away.  The true VEVs depend on scale
through wave function renormalization effects which are never large as can be
seen from the form of the renormalization group equations for the VEVs in the
Appendix A.

In this analysis, we shall also reject solutions based on fine tuning
considerations; however, our method differs somewhat from the usual one in
that it is incorporated in the solution routine described above.  The routine
is an iterative one which determines the convergence properties of the
solution.  Very slow convergence reflects an inherent fine tuning.  Therefore,
if the convergence is too slow, we will reject the solution.  Effectively we
are rejecting any solution which the computer cannot pinpoint within an
allotted number of iterations.

Given values for $A_0$, $m_0$, $m_{1/2}$,
$\tan\beta$, and ${\rm sign}(\mu)$, the solution routines search for the
values of $v(M_X)$, $m_{b,\tau}(M_X)$, $m_t(M_X)$, $B_0$, and $\mu_0$.
The process by which $B_0$ and $\mu_0$ are found was described above.  The
remaining three parameters are determined similarly.  The routine makes a
guess for $v(M_X)$, $m_{b,\tau}(M_X)$, and $m_t(M_X)$, then the full
renormalization group equations are evolved to $1$ GeV calculating
superparticle threshold masses in the process and minimizing the one loop
effective potential at $M_Z$.  The merits of the guess for $v(M_X)$,
$m_{b,\tau}(M_X)$, and $m_t(M_X)$ are assessed by comparing the resulting
values of $M_Z$, $m_\tau(1~{\rm GeV})$, and $m_b(1~{\rm GeV})$ with the
expected ones.  The process is iterated until the correct values are
achieved to within a tolerance of $1\%$.

By a ``solution,'' we will mean that a choice for the four free parameters
(plus a choice for the sign of $\mu$) at $M_X$ results in a complete
set of all parameters at $M_X$ (using the shooting routines) consistent
with electroweak breaking ($M_Z=91.17$ GeV), with equality of the bottom
and $\tau$ masses at $M_X$, and with low energy experiment.

Such a solution then yields a precise spectrum of sparticle masses.  This
spectrum can be further used to limit the allowable free parameter space
by subjecting the spectrum to experimental restrictions.

\section{ thresholds }
\label{ thresholds }

In the minimal low energy supergravity model
being considered, the super particle spectrum is no longer degenerate as
in the simple global supersymmetry model in which all the super particles
are given a common mass, $M_{SUSY}$.  In this simple case, one makes one
course correction in the renormalization group evolution at $M_{SUSY}$.
In the model with soft symmetry breaking, however, the nondegenerate spectrum
should lead to various course corrections each occuring at the super particle
mass thresholds.  To this end, the renormalization group $\beta$-functions
must be cast in a new form which makes the implementation of the thresholds
effects (albeit naive) evident (see Appendix B).  Since the
${\overline {\rm MS}}$ renormalization group
equations are mass independent, particle thresholds must be handled using
the decoupling theorem \cite{appel}, and each super particle mass
has associated with it a boundary between two effective theories.  Above
a particular mass threshold the associated particle is present in the
effective theory, below the threshold the particle is absent.

The simplest way to incorporate this is to (naively) treat the
thresholds as steps in the particle content of the renormalization
group $\beta$-functions \cite{habhem}.  This method is not always
entirely adequate.  For example, in the case of the $SU(2)$ gauge
coupling there will be scales in the integration process at which
there are effectively a half integer number of doublets using this
method.  For example, in the region $M_b<\Lambda<M_t$ between the mass
of the bottom and top quarks, the number of quark doublets is taken to
be two and a half using this method.  A similar situation occurs with
sparticle doublets.  We believe, nevertheless, that this method does
yield the correct, general behavior of the evolution.  It is a simple
means of implementing the smearing effects of the non-degenerate super
particle spectrum.  The determination of the spectrum of masses is
done without iteration as is common in other analyses.  Our method
deduces the physical masses by solving the equation
$m(\Lambda)=\Lambda$ for each superparticle in the process of evolving
from $M_X$ to $1$ GeV.  The usual iterative method requires several
runs to find a consistent solution.

\section{ analysis }
\label{ analysis }

The tremendous computing task involved in analyzing the full parameter space
of the soft symmetry breaking models, using the methods described as
designed, would be far too time consuming given the computing facilities
available to us.  Therefore, in the following analysis, some simplifications
will be made in the procedural method.  First, only the heaviest family of
quarks and leptons will have non-zero mass.  This will cut down on the
CPU time required for the solution routines to consistently find the running
mass values at $M_X$ since there are three instead of nine masses to determine
after this simplification.  Second, as stated previously, the
value of the strong coupling at $M_Z$ will be allowed to vary from its
central value of $.113$ by at most $10\%$.  This translates into a similar
error in the bottom quark mass.  Third, the allotted number of Runge-Kutta
steps, involved in numerically integrating the renormalization group
equations, will be cut down to $\sim 100$.

Our method involves four input parameters $A_0$, $m_0$, $m_{1/2}$,
$\tan\beta$ (and the ${\rm sign}(\mu)$).  The output is $B_0$,
$\mu_0$, $M_t$, $M_h$, and all the masses of the extra particles
associated with the MSSM.  Efficient use of CPU time required that we
proceed as follows.  For a given model, our initial exploration of the
parameter space was performed in a coarse grained fashion.
$\tan\beta$ was most commonly coarse grained as $2$, $5$, and $10$,
with only some runs involving higher values (e.g., $15$, $20$).  The
other three input parameters were varied in steps of $50$ and $100$
GeV.  Values larger than $\sim 500$ GeV were rarely ever used.  We
subsequently narrowed down on the allowed hyperenvelope by fine
graining around the edges of the expected region (based on the coarse
graining results).

Our raw data consists of those runs which satisfy the following two
criteria.  The first is consistency with electroweak breaking; that
is, the correct value of $M_Z$ is achieved from the minimization of
the one loop effective potential with $\mu^2(M_Z)>0$.  The second
criterion is no fine tuning, as implemented in our method (see Section
VII).  The solution routines employed return a numbered code
representing the convergence properties of the solution which we use
to screen the runs.

The raw data can then be progressively filtered based on at least
three physical constraints.  The first is cosmological.  Conserved
R-parity ($+1$ for particles and $-1$ for superparticles) requires the
existence of a stable lightest supersymmetric particle (LSP).
Astrophysical considerations indicate that the LSP must be neutral and
colorless.  Cosmological considerations based on the LSPs contribution
to the density of the universe indicate that it must have a mass less
than $\sim 200$ GeV \cite{arnnat,robros,lnz}.  Therefore, points in
parameter space that lead to LSPs other than neutralinos with masses
less than $\sim 200$ GeV are cut.  Second, flavor changing neutral
current bounds can be used to reject runs in which interfamily
splitting of squark and slepton masses is too large, although this is
seldom the case in the type of low energy supergravity models with
soft breaking universality that we consider.  Third, experimental
limits on the masses of the superparticles can also be used as another
criterion to reject some runs.

\section{ results }
\label{ results }

In the following, we discuss the results of our analysis of the four
classes of models listed in Table II and of a general, non-constrained
class.  In the general case, we discuss results based on all our runs
with no constraints on the soft parameters.  Because we have performed
a coarse grained study, using fine graining only for selected regions,
our results should only be considered qualitatively valid.  Based on
these, we hope to be able to ascertain the general trends in the data,
and make some general predictions about the feasibility of the models
considered.

Because the GUT inspired constraint Eq.~(\ref{mbeqmtau}) is enforced
in this analysis, the results will depend on the mass of the bottom
quark.  Namely, lower bottom quark masses require larger values of the
top quark mass to satisfy this relation, and higher bottom quark
masses require smaller values of the top quark mass.  For simplicity,
most results will be reported for the case $m_b(1~{\rm GeV})=6.00$
GeV, but lower mass ($5.70$ GeV) and higher mass ($6.33$ GeV) cases
were also studied.  These lower and higher values are within
experimental uncertainty.  The running value of $6.00$ GeV for
$m_b(1~{\rm GeV})$ corresponds to a physical bottom mass $M_b=4.85\pm
.15$ GeV, with the uncertainty coming from the error in the strong
coupling, as discussed above.

In Table III, we present the allowable ranges of the soft breaking
parameters for the general unrestricted case and for each of the four
subcases considered.  Since $B_0$ and $\mu_0$ are not input parameters
in our method, we must infer the results for the strict no-scale,
string inspired, and minimal SUGRA cases.  We do this by setting a
small tolerance on $|B_0|$ and $|\chi|$ of $25$ GeV.  In compiling
this table, we have discarded all runs with $M_t<120$ GeV based on
what we feel is a reasonable experimental lower bound for the top
quark mass.  We note from the table that $\mu_0$ is less than $\sim
200$ GeV in the first three cases, and only in the minimal SUGRA case
does it reach larger values.  The largest value of $\mu_0$ was $817$
GeV and occured for a run that does not correspond to any of the above
cases.  The values of the other soft parameters in this run and the
resulting spectrum are displayed in Table IV.

\begin{table}
\caption{Allowed Soft Parameter Ranges (in units of GeV).}
\begin{tabular}{lccccc}
Case            & $A_0$     & $m_0$     & $m_{1/2}$   & $\mu_0$
                & $B_0$       \\
\hline
General         & $(0,500)$ & $(0,800)$ & $(190,400)$ & $(33,817)$
                & $(-56,1012)$ \\
No-Scale        & $0$       & $0$       & $(190,262)$ & $(40,230)$
                & $(-60,130)$ \\
Strict No-Scale & $0$       & $0$       & $(210,240)$ & $(159,188)$
                & $|B_0|<25$  \\
String Inspired & $0$       & $(0,200)$ & $(200,250)$ & $(88,188)$
                & $|B_0|<25$  \\
Minimal SUGRA   & $(0,500)$ & $(0,300)$ & $(200,300)$ & $(159,467)$
                & $(-12,407)$ \\
\end{tabular}
\end{table}


\subsection{ General Case }
\label{ General Case }

Based on all runs and imposing no particular constraints on any soft
parameters, we find that $86<M_t<185$ GeV for the top quark and
$35<M_h<141$ GeV for Higgs boson.
Table IV contains three scenarios with input parameters that do not
conform to any of the four aforementioned cases.  Scenario (a) has
a very large value of $m_0$.  The sparticle masses are the largest we
have encounterd so far in our discussion.  One interesting feature
of this scenario is that the lighter stop is lighter than any of the
sleptons.  Scenario (b) has a gluino of mass $908$ GeV and a top quark of
mass $184$ GeV.  The Higgs boson mass is one of the heaviest of all the
scenarios discussed at $132$ GeV.  Both (a) and (b) have very large
neutralino masses $\sim 170$ GeV.  The spectrum in (c) represents another
scenario in the general case, but with a larger value of $\tan\beta$.

\begin{table}
\caption{Sample Particle Spectra for General Models (in units of GeV).}
\begin{tabular}{lccccc}
                          & (a)          & (b)       & (c)    \\
\hline
$A_0$                     & $0$          & $200$     & $0$     \\
$m_0$                     & $800$        & $500$     & $100$     \\
$m_{1/2}$                 & $400$        & $400$     & $300$   \\
$\tan\beta$               & $2$          & $5$       & $15$    \\
${\rm sign}(\mu)$         & $+$          & $+$       & $+$     \\
$B_0$                     & $638$        & $326$     & $44$    \\
$\mu_0$                   & $817$        & $687$     & $301$   \\
                          &              &           &         \\
$M_t$                     & $175$        & $184$     & $155$   \\
${\tilde g}$                    & $913$        & $913$     & $692$      \\
                          &              &           &         \\
${\tilde u}_1, {\tilde u}_2$    & $1112,1137$  & $926,956$ & $604,629$  \\
${\tilde t}_1, {\tilde t}_2$    & $696,993$    & $626,869$ & $476,636$  \\
${\tilde d}_1, {\tilde d}_2$    & $1109,1139$  & $922,959$ & $602,634$  \\
${\tilde b}_1, {\tilde b}_2$    & $966,1109$   & $831,923$ & $573,611$  \\
${\tilde e}_1, {\tilde e}_2$    & $815,847$    & $525,573$  & $159,236$  \\
${\tilde\tau}_1,{\tilde\tau}_2$ & $815,847$    & $521,573$  & $143,241$  \\
${\tilde\nu}_e$                 & $844$        & $568$     & $223$      \\
${\tilde\nu}_\tau$              & $844$        & $567$     & $221$      \\
                          &              &           &         \\
${\tilde C}_1$                  & $324$        & $323$     & $222$      \\
${\tilde C}_2$                  & $724$        & $604$     & $358$      \\
                          &              &           &         \\
${\tilde N}_1$                  & $171$        & $171$     & $125$       \\
${\tilde N}_2$                  & $324$        & $324$     & $223$      \\
${\tilde N}_3$                  & $-712$       & $-590$    & $-326$     \\
${\tilde N}_4$                  & $726$        & $604$     & $358$      \\
                          &              &           &         \\
$h$                             & $118$        & $132$     & $112$      \\
$H$                             & $1149$       & $755$     & $178$      \\
$A$                             & $1147$       & $755$     & $177$      \\
$H^\pm$                         & $1150$       & $759$     & $194$      \\
\end{tabular}
\end{table}

\subsection{ No-scale Case }
\label{ No-scale Case }

In all cases considered, the mass of the LSP, when it is a neutralino,
is observed to be correlated with the value of $m_{1/2}$.  Therefore,
we find that $m_{1/2}$ cannot be taken too large ($\alt 400$ GeV).  In
the no-scale case, we present plots of $M_t$ vs. $m_{1/2}$ for three
values of $m_b(1~{\rm GeV})$ and containing all points satisfying the
various criteria outlined in Section IX.  Figure 1 is such a plot for
$m_b(1~{\rm GeV})=6.00$ GeV.  The ``right edge'' of the envelope is
defined by points whose neutralino LSPs are just slightly heavier than
the lightest charged superparticle (usually a
${\tilde\tau}_{\scriptscriptstyle R}$ for us).  The ``left edge''
defines the threshold of consistent electroweak breaking.  The top and
bottom edges are set by the requirement that Eq.~(\ref{mbeqmtau})
hold.  The range of $\tan\beta$ considered leads to definite lower and
upper bounds on $M_t$ \cite{summer91,kln,anan,giveon}.  In Figs.~2 and
3, we display similar plots with lower ($m_b(1~{\rm GeV})=5.70$ GeV)
and higher ($m_b(1~{\rm GeV})=6.33$ GeV) bottom quark masses.  The
previously noted dependence of the data on $M_b$ is evident from these
figures.

{}From these, we note that the available range of $m_{1/2}$ decreases
with increasing $M_b$.  Fig.~1 indicates that $190~{\rm GeV}\alt
m_{1/2}\alt 265~{\rm GeV}$.  We can draw no conclusions, however,
about the value of $\tan\beta$.  Although, we found that values of
$\tan\beta$ larger than 18 never led to solutions in the no-scale
case.  We can conclude from the $m_b(1~{\rm GeV})=6.00$ GeV case that
$85 \alt M_t \alt 132$ GeV.  From similar plots involving $M_h$, we
conclude that $35 \alt M_h \alt 110$ GeV.  In Fig.~4, we display the
familiar dependence of $M_t$ on $\tan\beta$ for a particular value of
$m_{1/2}$ in the allowed range.

The official experimental lower bound on the top quark mass is $108$
GeV.  The figures indicate that the top quark cannot have a mass
greater than $\sim 132$ GeV in this model, if $m_b(1~{\rm GeV})=6.00$
GeV.  This upper bound is raised to $\sim 160$ GeV, if $m_b(1~{\rm
GeV})=5.70$ GeV, and the model is ruled out, if $m_b(1~{\rm
GeV})=6.33$ GeV.

Figure 5 is similar to Fig.~1, but we have chosen the sign of $\mu$
negative.  The allowed region is displaced down with respect to the
positive $\mu$ case with the upper bound on $M_t$ now $113$ GeV and
very close to the experimental limit.

\begin{table}
\caption{Sample Particle Spectra for  No-Scale Models (in units of GeV).}
\begin{tabular}{lccccc}
                          & (a)          & (b)       & (c)    \\
\hline
$A_0$                     & $0$          & $0$       & $0$     \\
$m_0$                     & $0$          & $0$       & $0$     \\
$m_{1/2}$                 & $254$        & $240$     & $240$   \\
$\tan\beta$               & $5$          & $3$       & $10$    \\
${\rm sign}(\mu)$         & $+$          & $+$       & $+$     \\
$B_0$                     & $29$         & $66$      & $-12$   \\
$\mu_0$                   & $203$        & $198$     & $166$   \\
                          &              &           &         \\
$M_t$                     & $131$        & $126$     & $124$   \\
${\tilde g}$                    & $589$        & $558$     & $558$      \\
                          &              &           &         \\
${\tilde u}_1, {\tilde u}_2$    & $508,530$    & $482,502$ & $482,501$  \\
${\tilde t}_1, {\tilde t}_2$    & $408,559$    & $381,535$ & $392,532$  \\
${\tilde d}_1, {\tilde d}_2$    & $507,535$    & $481,507$ & $481,507$  \\
${\tilde b}_1, {\tilde b}_2$    & $507,508$    & $476,486$ & $472,491$  \\
${\tilde e}_1, {\tilde e}_2$    & $107,183$    & $101,173$  & $102,174$  \\
${\tilde\tau}_1,{\tilde\tau}_2$ & $105,184$    & $100,173$  & $97,176$  \\
${\tilde\nu}_e$                 & $166$        & $157$     & $155$      \\
${\tilde\nu}_\tau$              & $166$        & $157$     & $155$      \\
                          &              &           &         \\
${\tilde C}_1$                  & $164$        & $150$     & $145$      \\
${\tilde C}_2$                  & $296$        & $293$     & $267$      \\
                          &              &           &         \\
${\tilde N}_1$                  & $100$        & $90$      & $92$       \\
${\tilde N}_2$                  & $170$        & $157$     & $152$      \\
${\tilde N}_3$                  & $-246$       & $-240$    & $-208$     \\
${\tilde N}_4$                  & $298$        & $296$     & $268$      \\
$h$                             & $98$         & $88$      & $99$       \\
$H$                             & $278$        & $290$     & $203$      \\
$A$                             & $276$        & $285$     & $202$      \\
$H^\pm$                         & $287$        & $296$     & $217$      \\
\end{tabular}
\end{table}

Table V displays various spectra of superparticle masses in the no-scale
case.  In spectrum (a), we present a scenario in which
$m_{1/2}=254$ GeV and $\tan\beta=5$.  In this particular scenario, the top
quark mass is $131$ GeV and slightly below is the mass of the light Higgs
boson at $98$ GeV.  A qualitative feature of most spectra is that the
sleptons are lighter than the squarks.  Another feature is that one
of the stops is lighter than all other squarks.  In this table, as in
similar ones to follow, we do not include the second family sfermion
masses since they are generally degenerate with those of the first family.
Also, we have singled out the flippino by associating with it a negative
mass.  The input parameters of spectra (b) and (c) differ only in the
respective values of $\tan\beta$, $3$ and $10$.  The spectra are identical in
almost every respect except for the splitting in the stop masses.  The
stop mass splitting in spectrum (b) is $154$ GeV and $140$ GeV in spectrum
(c) as expected since the product $\mu\cot\beta$ is bigger in case (b).
Spectrum (a) represents a shifted version of (b) or (c).


\subsection{ Strict No-scale Case }
\label{ Strict No-scale Case }

The results of the strict no-scale case must be interpolated from the
no-scale results, because $B_0$ is not an input parameter in our procedural
method.  Therefore, we plot in Fig.~6 $M_t$ vs. $B_0$ and deduce the $M_t$
bounds from slicing the data along $B_0=0$.  Admittedly, the relatively
small number of points makes the perimeter of the region unclear in some
areas.  We get approximately $90 \alt M_t \alt 127$ GeV.  For $M_h$, we
find $57 \alt M_h \alt 110$ GeV with an uncertainty in the lower bound of
$\sim 10$ GeV due to lack of definition in the lower end of the envelope.
Inspection of the data indicates that $3.5 \alt \tan\beta \alt 9$.  Thus,
it appears that $\tan\beta$ cannot be too small or too large to accommodate
the strict no-scale case.  This case is a special case of both the
no-scale and string inspired cases.

Finally, in Table VI, we display spectra of superparticle masses for three
representative strict no-scale scenarios with various $\tan\beta$ and
$m_{1/2}$.  Although, as evident from the table, none of these scenarios
have $B_0$ strictly equal to zero, they do satisfy $|B_0|<25$ GeV, which
is the tolerance we have settled upon.  The LSPs are neutralinos in these
three scenarios and have masses that increase with increasing $m_{1/2}$.
The top quark masses are comparable at $\sim 126$ GeV and just above the
experimental limit, and the Higgs bosons masses are also approximately
equal $\sim 99$ GeV.  As in the no-scale case, we observe that increasing
the value of $m_{1/2}$ has the effect of shifting the squark and
slepton spectra up.

\begin{table}
\caption{Sample Particle Spectra for  Strict No-Scale Models
         (in units of GeV).}
\begin{tabular}{lccccc}
                          & (a)          & (b)       & (c)    \\
\hline
$A_0$                     & $0$          & $0$       & $0$     \\
$m_0$                     & $0$          & $0$       & $0$     \\
$m_{1/2}$                 & $240$        & $230$     & $210$   \\
$\tan\beta$               & $8.3$        & $10$      & $5$    \\
${\rm sign}(\mu)$         & $+$          & $+$       & $+$     \\
$B_0$                     & $-4$         & $-11$     & $18$   \\
$\mu_0$                   & $171$        & $159$     & $159$   \\
                          &              &           &         \\
$M_t$                     & $126$        & $124$     & $128$   \\
${\tilde g}$                    & $558$        & $536$     & $491$      \\
                          &              &           &         \\
${\tilde u}_1, {\tilde u}_2$    & $482,501$    & $462,481$ & $424,441$  \\
${\tilde t}_1, {\tilde t}_2$    & $390,533$    & $373,514$ & $331,480$  \\
${\tilde d}_1, {\tilde d}_2$    & $481,507$    & $462,487$ & $423,447$  \\
${\tilde b}_1, {\tilde b}_2$    & $473,490$    & $453,472$ & $418,430$  \\
${\tilde e}_1, {\tilde e}_2$    & $102,174$    & $99,167$  & $92,154$  \\
${\tilde\tau}_1,{\tilde\tau}_2$ & $98,175$     & $93,169$  & $90,154$  \\
${\tilde\nu}_e$                 & $156$        & $148$     & $133$      \\
${\tilde\nu}_\tau$              & $155$        & $148$     & $133$      \\
                          &              &           &         \\
${\tilde C}_1$                  & $147$        & $137$     & $122$      \\
${\tilde C}_2$                  & $270$        & $259$     & $253$      \\
                          &              &           &         \\
${\tilde N}_1$                  & $92$         & $87$      & $76$       \\
${\tilde N}_2$                  & $153$        & $144$     & $131$      \\
${\tilde N}_3$                  & $-213$       & $-200$    & $-197$     \\
${\tilde N}_4$                  & $271$        & $260$     & $255$      \\
                          &              &           &         \\
$h$                             & $99$         & $99$      & $96$      \\
$H$                             & $222$        & $194$     & $223$      \\
$A$                             & $221$        & $193$     & $221$      \\
$H^\pm$                         & $235$        & $209$     & $234$      \\
\end{tabular}
\end{table}

\begin{table}
\caption{Sample Particle Spectra for String Inspired Models (in units of GeV).}
\begin{tabular}{lccccc}
                          & (a)          & (b)       & (c)    \\
\hline
$A_0$                     & $0$          & $0$       & $0$     \\
$m_0$                     & $100$        & $100$     & $200$     \\
$m_{1/2}$                 & $200$        & $250$     & $200$   \\
$\tan\beta$               & $10$         & $10$      & $10$    \\
${\rm sign}(\mu)$         & $+$          & $+$       & $+$     \\
$B_0$                     & $-4$         & $-6$      & $20$   \\
$\mu_0$                   & $132$        & $168$     & $88$   \\
                          &              &           &         \\
$M_t$                     & $124$        & $125$     & $123$   \\
${\tilde g}$                    & $469$        & $580$     & $469$      \\
                          &              &           &         \\
${\tilde u}_1, {\tilde u}_2$    & $416,431$    & $510,530$ & $449,464$  \\
${\tilde t}_1, {\tilde t}_2$    & $330,467$    & $417,556$ & $361,489$  \\
${\tilde d}_1, {\tilde d}_2$    & $416,438$    & $509,536$ & $449,470$  \\
${\tilde b}_1, {\tilde b}_2$    & $407,424$    & $509,509$ & $440,453$  \\
${\tilde e}_1, {\tilde e}_2$    & $134,178$    & $145,206$ & $219,248$  \\
${\tilde\tau}_1,{\tilde\tau}_2$ & $129,179$    & $141,208$ & $216,248$  \\
${\tilde\nu}_e$                 & $160$        & $191$     & $235$      \\
${\tilde\nu}_\tau$              & $159$        & $190$     & $234$      \\
                          &              &           &         \\
${\tilde C}_1$                  & $109$        & $150$     & $77$      \\
${\tilde C}_2$                  & $231$        & $272$     & $214$      \\
                          &              &           &         \\
${\tilde N}_1$                  & $71$         & $96$      & $53$       \\
${\tilde N}_2$                  & $119$        & $157$     & $105$      \\
${\tilde N}_3$                  & $-170$       & $-209$    & $-120$     \\
${\tilde N}_4$                  & $232$        & $273$     & $214$      \\
                          &              &           &         \\
$h$                             & $98$         & $100$     & $98$       \\
$H$                             & $189$        & $226$     & $230$      \\
$A$                             & $188$        & $225$     & $229$      \\
$H^\pm$                         & $204$        & $238$     & $242$      \\
\end{tabular}
\end{table}


\subsection{ String Inspired Case }
\label{ String Inspired Case }

As in the strict no-scale, our results for the string inspired case
necessitates interpolating $A_0=0$ data to $B_0=0$.  Once again we find
the perimeter of the allowed region is not well defined everywhere.  Hence,
our results are only qualitative.  Figure 7 is similar to Fig.~6, but in
this case we only fix $A_0=0$.
Since the strict no-scale case is a special case of the string inspired one,
the $127$ GeV top quark mass upper bound is not expected to
decrease but rather to increase in this case.
Slicing along $B_0=0$ yields
$85 \alt M_t \alt 140$ GeV.  Similarly, for the light Higgs we get
$57 \alt M_h \alt 113$ GeV.  The data indicates in this case, as in the
strict no-scale case, that there is a lower bound on $\tan\beta$ of $\sim 3$.

In Table VII, we collect some spectra in the string inspired case.  As in
the strict no-scale case, we have settled on a tolerance of $|B_0|<25$ GeV for
the scenarios considered.  Spectrum (a) with input parameters $A_0=0$,
$m_0=100$ GeV, $m_{1/2}=200$ GeV, and $\tan\beta=10$ has $B_0=-4$ GeV.
The strict $B_0=0$ run has a value of $\tan\beta$ between $5$ and $10$, but
the qualitativee features should be very similar to the scenario we
present with $\tan\beta=10$.  Scenario (b) is similar to scenario (a)
except it has a larger value of $m_{1/2}=250$ GeV.  As noted before, this
leads to an upward shift in the squark and slepton spectrum of (b).  The
upward shift is $23\%$ in the squark masses and $15\%$ in the slepton masses.
Scenario (c) is similar to scenario (a) except it has a larger value of
$m_0=200$ GeV.  This shifts up the squark and slepton masses of (c) with
respect to (a).  In this case, the squarks shift up by only $7\%$ and the
sleptons shift up by $49\%$.  Therefore, spectra (b) and (c) have the
curious property that the squarks of (b) are heavier than those of (c), but
the sleptons of (b) are lighter than those of (c).  Lastly, we
observe that the lighter selectron and stau of all spectra discussed so far
display a comparatively large mass gap from the rest of their respective
slepton partners, however in scenario (c) this mass gap is comparatively
small; and all sleptons are almost degenerate.


\subsection{ Minimal SUGRA Case }
\label{ Minimal SUGRA Case }

\begin{table}
\caption{Sample Particle Spectra for  Minimal SUGRA Models (in units of GeV).}
\begin{tabular}{lccccc}
                          & (a)          & (b)       & (c)    \\
\hline
$A_0$                     & $400$        & $500$     & $400$     \\
$m_0$                     & $200$        & $200$     & $300$     \\
$m_{1/2}$                 & $300$        & $300$     & $200$   \\
$\tan\beta$               & $10$         & $5$       & $10$    \\
${\rm sign}(\mu)$         & $+$          & $+$       & $+$     \\
$B_0$                     & $212$        & $288$     & $115$   \\
$\mu_0$                   & $404$        & $436$     & $158$   \\
                          &              &           &         \\
$M_t$                     & $164$        & $163$     & $124$   \\
${\tilde g}$                    & $692$        & $692$     & $469$      \\
                          &              &           &         \\
${\tilde u}_1, {\tilde u}_2$    & $627,652$    & $627,652$ & $500,513$  \\
${\tilde t}_1, {\tilde t}_2$    & $415,643$    & $389,644$ & $330,529$  \\
${\tilde d}_1, {\tilde d}_2$    & $625,656$    & $625,656$ & $500,519$  \\
${\tilde b}_1, {\tilde b}_2$    & $573,630$    & $574,627$ & $463,502$  \\
${\tilde e}_1, {\tilde e}_2$    & $235,293$    & $235,292$ & $313,334$  \\
${\tilde\tau}_1,{\tilde\tau}_2$ & $221,295$    & $230,294$ & $306,333$  \\
${\tilde\nu}_e$                 & $282$        & $282$     & $324$      \\
${\tilde\nu}_\tau$              & $279$        & $281$     & $322$      \\
                          &              &           &         \\
${\tilde C}_1$                  & $236$        & $236$     & $123$      \\
${\tilde C}_2$                  & $438$        & $471$     & $247$      \\
                          &              &           &         \\
${\tilde N}_1$                  & $127$        & $126$     & $76$       \\
${\tilde N}_2$                  & $236$        & $236$     & $129$      \\
${\tilde N}_3$                  & $-421$       & $-454$    & $-200$     \\
${\tilde N}_4$                  & $438$        & $472$     & $248$      \\
                          &              &           &         \\
$h$                             & $119$        & $118$     & $101$      \\
$H$                             & $371$        & $491$     & $325$      \\
$A$                             & $371$        & $490$     & $325$      \\
$H^\pm$                         & $379$        & $497$     & $334$      \\
\end{tabular}
\end{table}

We now consider models for which the relation $B_0=A_0-m_0$ among
the soft couplings holds.  We find it convenient to define
$\chi\equiv A_0-B_0-m_0$, which should equal zero in this case.
In Fig.~8, we plot $M_t$ vs. $\chi$.  Taking a slice along $\chi=0$ of the
region depicted in this plot, we find that $82 \alt M_t \alt 185$ GeV.  A
similar plot was used to arrive at bounds for the light Higgs boson
$57 \alt M_h \alt 139$ GeV.  Our analysis indicates that there is no preferred
values of $\tan\beta$ in this case.

Table VIII includes the spectra for some particular scenarios in the
minimal SUGRA case.  The three spectra appearing in this table have the
interesting feature that the slepton masses are larger than they have
been in all cases discussed so far.  In fact the slepton masses have
values comparable to the lighter stop in (a), (b), and (c).  These three
scenarios each have relatively large, non-zero values of $A_0$.  As expected,
scenario (b) has the larger value of $\mu\cot\beta$ and therefore has the
larger splitting in stop masses when compared to (a).  Indeed, this is the
only significant difference in these two scenarios.  Scenario (c) in the
minimal SUGRA case has a fairly degenerate slepton sector like that of
(c) in Table VII.  Once again we observe the curious feature when comparing
(a) and (c) that the squarks of (a) are heavier than those of (c), but
the sleptons of (a) are lighter than those of (c).  It appears that
the run with the larger value of $m_0$ has the heavier sleptons and that
with the larger value of $m_{1/2}$ has the heavier squarks.
Scenarios (a) and (b) are interesting because they have large
values of $\mu\tan\beta$ which leads to large splittings in the
bottom squark and stau masses.  This represents an example where neglecting
the bottom or $\tau$ Yukawa can radically alter the spectrum.
The value of $M_t$ in (a) and (b) is $\sim 163$ GeV, however in (c) it
is $124$ GeV.  We find based on studying other similar runs that the input
parameter accounting mostly for this large difference is the value of
$m_{1/2}$.  Both (a) and (b) have $m_{1/2}=300$ GeV whereas (c) has
$m_{1/2}=200$ GeV.  We observe also that (a) and (b) have larger (comparable)
values of $\mu_0$ and $B_0$ whereas (c) has a $\sim 50\%$ smaller values of
these input parameters.



\section{ conclusions }
\label{ conclusions }

Minimal low energy supergravity models were considered.  They have the
appealing feature that the electro-weak symmetry is radiatively broken
for certain ranges of the soft breaking parameters and of the top
quark mass.  The study of specific models, with some soft parameter
fixed or related, resulted in upper bounds for the top quark mass.
No-scale models in which only gaugino masses provide global
supersymmetry breaking yield top quarks with masses less than $\sim
127$ GeV.  The results are sensitive to the value of the bottom quark
mass.  Lower bottom quark masses, within the experimental uncertainty,
lead to higher top quark upper bounds.  In these models, the ratio of
vacuum expectation values of the two Higgs fields is expected to be
larger than $\sim 70^\circ$.

Although the perimeter of the allowed regions were often fuzzy, we
could, nevertheless, draw some general conclusions from our results.
For all our runs, with no restrictions on the soft terms, we find for
the top quark $M_t \alt 185$ GeV and for the light Higgs boson $M_h
\alt 141$ GeV .

\section{ acknowledgments}
\label{  acknowledgments}

The authors wish to thank Drs Steve Martin and Haukur Arason for
useful discussions.  For their kind hospitality during the completion
of this work, (PR) wishes to thank the Aspen Center for Physics and
(DJC) would like to thank Prof. Iba\~nez and his group in Madrid. This
work has been supported in part by the United States Department of
Energy under contract D.E. FG05-86ER-40272.

\appendix
\section{ THE SUSY $\beta$-FUNCTIONS }
\label{ susy-beta }

Using some of the notation of Falck \cite{falck}, the superpotential and
soft symmetry breaking potential are as follows:
\widetext
\begin{eqnarray}
   W &=&  {\hat{\overline u}} {\bf Y}_u^{} {\hat \Phi}_u^{} {\hat Q}
        + {\hat{\overline d}} {\bf Y}_d^{} {\hat \Phi}_d^{} {\hat Q}
        + {\hat{\overline e}} {\bf Y}_e^{} {\hat \Phi}_d^{} {\hat L}
        + \mu {\hat \Phi}_u {\hat \Phi}_d + h.c. \ , \\
   V_{soft} &=&  m_{\Phi_u}^2 \Phi_u^\dagger \Phi_u^{}
               + m_{\Phi_d}^2 \Phi_d^\dagger \Phi_d^{}
               + B \mu ( \Phi_u \Phi_d + h.c. ) \nonumber \\
   & &\mbox{}  + \sum_i (\;
                 m_{{\tilde Q}_i}^2 {\tilde Q}_i^\dagger {\tilde Q}_i
               + m_{{\tilde L}_i}^2 {\tilde L}_i^\dagger {\tilde L}_i
               + m_{{\tilde{\overline u}}_i}^2 {\tilde{\overline u}}_i^\dagger
                             {\tilde{\overline u}}_i
               + m_{{\tilde{\overline d}}_i}^2 {\tilde{\overline d}}_i^\dagger
                             {\tilde{\overline d}}_i
               + m_{{\tilde{\overline e}}_i}^2 {\tilde{\overline e}}_i^\dagger
                             {\tilde{\overline e}}_i \; ) \nonumber \\
    & &\mbox{} + \sum_{i,j}\Bigl(\;
                 A_u^{ij}Y_u^{ij}{\tilde{\overline u}}_i\Phi_u{\tilde Q}_j
               + A_d^{ij}Y_d^{ij}{\tilde{\overline d}}_i\Phi_d{\tilde Q}_j
               + A_e^{ij}Y_e^{ij}{\tilde{\overline e}}_i\Phi_d{\tilde L}_j
               + h.c.
                 \;\Bigr) \ , \\
   V_{gaugino} &=& {1\over2} \sum_{l=1}^3
                         M_l \lambda_l \lambda_l + h.c. \ .
\end{eqnarray}
\narrowtext
Various $\sigma_2$'s have been omitted and a sum over the number of
families is implied in the squark and slepton mass terms.
Also, hats imply superfields and tildes the superpartners of the given fields.

We start with the gauge couplings
\begin{equation}
   {dg_l^{}\over dt} = - {1\over16\pi^2} b_l^{} g_l^3
    + {g_l^3\over (16\pi^2)^2} \Bigl[ \sum_k b_{lk}^{} g_k^2
   - {\rm Tr}\{ C_{lu}{\bf Y}_u^\dagger{\bf Y}_u^{}
   + C_{ld}{\bf Y}_d^\dagger{\bf Y}_d^{}
   + C_{le}{\bf Y}_e^\dagger{\bf Y}_e^{} \} \Bigr]
\end{equation}
where $t=\ln\Lambda$ and $l=1,\ 2,\ 3$, corresponding to gauge group
$\rm SU(3)_C \times SU(2)_L \times U(1)_Y$ of the Standard Model.

In the Yukawa sector the $\beta$-functions are
\begin{equation}
   {d{\bf Y}_{u,d,e}^{} \over dt} = {\bf Y}_{u,d,e}^{} (
   {1\over16\pi^2} {\bf \beta}_{u,d,e}^{(1)} + {1\over(16\pi^2)^2}
   {\bf \beta}_{u,d,e}^{(2)} )\ .
\end{equation}

The evolution of the vacuum expectation values of the Higgs's is given by
\begin{equation}
   {d\ln v_{\Phi_u,\Phi_d} \over dt} = {1\over16\pi^2}
   \gamma^{(1)}_{\Phi_u,\Phi_d} + {1\over(16\pi^2)^2}
   \gamma^{(2)}_{\Phi_u,\Phi_d} \ .
\label{vevrge}
\end{equation}

\subsection{ The one loop SUSY $\beta$-functions }

The various one loop coefficients for the gauge couplings are defined to be
\begin{equation}
\left\{ \begin{array}{ccr}
   b_1 &=& - {3\over5} - 2 n_g \ , \nonumber \\
   b_2 &=& 5 - 2 n_g \ , \\
   b_3 &=& 9 - 2 n_g \ , \nonumber
\end{array} \right.
\end{equation}
with $n_g={1\over2}n_{fl}$.

In the following, we list the one loop contributions for the parameters of the
superpotential.
\begin{equation}
   {d\ln\mu\over dt} = {1\over16\pi^2}[\; {\rm Tr}\{
              3 {\bf Y}_u^\dagger{\bf Y}_u^{} + 3 {\bf Y}_d^\dagger{\bf Y}_d^{}
                        + {\bf Y}_e^\dagger{\bf Y}_e^{} \}
                         - 3 ( {1\over5}g_1^2 + g_2^2 ) \;] \ .
\end{equation}

The one-loop contributions for the Yukwas are given by
\begin{eqnarray}
   {\bf \beta}_u^{(1)} = 3&& {\bf Y}_u^\dagger {\bf Y}_u^{}
                   + {\bf Y}_d^\dagger {\bf Y}_d^{}
                     + 3 {\rm Tr}\{{\bf Y}_u^\dagger{\bf Y}_u^{}\}
   - ( {13\over15}g_1^2 + 3g_2^2 + {16\over3}g_3^2 ) \ ,
   \label{1lpubf} \\
   {\bf \beta}_d^{(1)} = 3&& {\bf Y}_d^\dagger {\bf Y}_d^{}
          + {\bf Y}_u^\dagger {\bf Y}_u^{}
          + {\rm Tr}\{3{\bf Y}_d^\dagger{\bf Y}_d^{}
          +{\bf Y}_e^\dagger{\bf Y}_e^{}\}
   - ( {7\over15}g_1^2 + 3g_2^2 + {16\over3}g_3^2 ) \ ,
   \label{1lpdbf} \\
   {\bf \beta}_e^{(1)} = 3&& {\bf Y}_e^\dagger {\bf Y}_e^{}
       + {\rm Tr}\{3{\bf Y}_d^\dagger{\bf Y}_d^{}
          +{\bf Y}_e^\dagger{\bf Y}_e^{}\}
   - ( {9\over5}g_1^2 + 3g_2^2 ) \ .
   \label{1lpebf}
\end{eqnarray}

The one-loop contributions for the running VEVs are
\begin{eqnarray}
   \gamma^{(1)}_{\Phi_u} &=& {3\over4}( {1\over5}g_1^2 + g_2^2 )
                            - 3{\rm Tr}\{{\bf Y}_u^\dagger{\bf Y}_u^{}\}\ ,\\
   \gamma^{(1)}_{\Phi_d} &=& {3\over4}( {1\over5}g_1^2 + g_2^2 )
                            - 3{\rm Tr}\{{\bf Y}_d^\dagger{\bf Y}_d^{}\}
                      - {\rm Tr}\{{\bf Y}_e^\dagger{\bf Y}_e^{}\} \ .
\end{eqnarray}

\widetext
The soft symmetry breaking terms are known to us only to one loop
\begin{eqnarray}
   {d A_e^{ij}\over dt} &=& {1\over16\pi^2} [\; 4({\bf Y}_e^{}
        {\bf Y}_e^\dagger)^{ik} A_e^{kj}
   {Y_e^{kj}\over Y_e^{ij}} + 5A_e^{ik}{Y_e^{ik}\over Y_e^{ij}}
   ({\bf Y}_e^\dagger{\bf Y}_e^{})^{kj} - 3{A_e^{ij}\over Y_e^{ij}}
      ({\bf Y}_e^{}{\bf Y}_e^\dagger{\bf Y}_e^{})^{ij} \nonumber \\
   &&\mbox{} + 2( A_e^{km}\vert Y_e^{km}\vert^2 + 3A_d^{km}
   \vert Y_d^{km}\vert^2 ) - 6( {3\over5}g_1^2M_1 + g_2^2M_2 )\;]\ ,\\
   {d A_d^{ij}\over dt} &=& {1\over16\pi^2} [\; 4({\bf Y}_d^{}
          {\bf Y}_d^\dagger)^{ik} A_d^{kj}
   {Y_d^{kj}\over Y_d^{ij}} + 5A_d^{ik}{Y_d^{ik}\over Y_d^{ij}}
   ({\bf Y}_d^\dagger{\bf Y}_d^{})^{kj}
       - 3{A_d^{ij}\over Y_d^{ij}}
        ({\bf Y}_d^{}{\bf Y}_d^\dagger{\bf Y}_d^{})^{ij} \nonumber \\
   &&\mbox{}+ (A_d^{ik}-A_d^{ij})({\bf Y}_u^\dagger{\bf Y}_u^{})^{kj}
          {Y_d^{ik}\over Y_d^{ij}}
      + 2({\bf Y}_d^{}{\bf Y}_u^\dagger)^{ik} A_u^{kj}
          {Y_u^{kj}\over Y_d^{ij}}
   + 2( A_e^{km}\vert Y_e^{km}\vert^2 \nonumber \\
    &&\mbox{}+ 3A_d^{km}
   \vert Y_d^{km}\vert^2 )
  - {14\over15}g_1^2M_1 - 6g_2^2M_2 -
   {32\over3}g_3^2M_3 \;] \ , \\
   {d A_u^{ij}\over dt} &=& {1\over16\pi^2} [\; 4({\bf Y}_u^{}
          {\bf Y}_u^\dagger)^{ik} A_u^{kj}
   {Y_u^{kj}\over Y_u^{ij}} + 5A_u^{ik}{Y_u^{ik}\over Y_u^{ij}}
   ({\bf Y}_u^\dagger{\bf Y}_u^{})^{kj} - 3{A_u^{ij}\over Y_u^{ij}}
       ({\bf Y}_u^{}{\bf Y}_u^\dagger{\bf Y}_u^{})^{ij} \nonumber \\
   &&\mbox{}   + (A_u^{ik}-A_u^{ij})({\bf Y}_d^\dagger{\bf Y}_d^{})^{kj}
         {Y_u^{ik}\over Y_u^{ij}}
     + 2({\bf Y}_u^{}{\bf Y}_d^\dagger)^{ik} A_d^{kj}
            {Y_d^{kj}\over Y_u^{ij}}
   + 6A_u^{km}\vert Y_u^{km}\vert^2 \nonumber \\
   &&\mbox{}- {26\over15}g_1^2M_1 - 6g_2^2M_2 - {32\over3}g_3^2M_3 \;] \ , \\
   {dm_{\Phi_u}^2 \over dt} &=& {1\over8\pi^2}[\; \sum_{i,j}
   3\vert Y_u^{ji}\vert^2 ( m_{\Phi_u}^2 + m_{Q_i}^2 + m_{u_j}^2 +
   \vert A_u^{ji}\vert^2 )
 + {3\over10}g_1^2{\rm Tr}\{Ym^2\} - {3\over5}g_1^2M_1^2 \nonumber \\
   &&\mbox{}- 3g_2^2M_2^2 \;] \ , \label{appeq1} \\
   {dm_{\Phi_d}^2 \over dt} &=& {1\over8\pi^2}[\; \sum_{i,j} \Bigl(
   \vert Y_e^{ji}\vert^2 ( m_{\Phi_d}^2 + m_{L_i}^2 + m_{e_j}^2 +
   \vert A_e^{ji}\vert^2 )
   + 3\vert Y_d^{ji}\vert^2 ( m_{\Phi_d}^2 + m_{Q_i}^2 + m_{d_j}^2 \nonumber \\
   &&\mbox{} + \vert A_d^{ji}\vert^2 ) \Bigr)
     - {3\over10}g_1^2{\rm Tr}\{Ym^2\} - {3\over5}g_1^2M_1^2
   - 3g_2^2M_2^2 \;] \ , \label{appeq2} \\
   {dm_{e_i}^2 \over dt} &=& {1\over8\pi^2}[\; \sum_{j}
   2\vert Y_e^{ij}\vert^2 ( m_{\Phi_d}^2 + m_{e_i}^2 + m_{L_j}^2 +
   \vert A_e^{ij}\vert^2 )
   + {3\over5}g_1^2{\rm Tr}\{Ym^2\} - {12\over5}g_1^2M_1^2 \;]
   \ , \label{appeq3} \\
   {dm_{L_i}^2 \over dt} &=& {1\over8\pi^2}[\; \sum_{j}
   \vert Y_e^{ji}\vert^2 ( m_{\Phi_d}^2 + m_{L_i}^2 + m_{e_j}^2 +
   \vert A_e^{ji}\vert^2 ) - {3\over10}g_1^2{\rm Tr}\{Ym^2\} \nonumber \\
   &&\mbox{} - {3\over5}g_1^2M_1^2
   - 3g_2^2M_2^2 \;] \ , \label{appeq4} \\
   {dm_{d_i}^2 \over dt} &=& {1\over8\pi^2}[\; \sum_{j}
   2\vert Y_d^{ij}\vert^2 ( m_{\Phi_d}^2 + m_{d_i}^2 + m_{Q_j}^2 +
   \vert A_d^{ij}\vert^2 )
   + {1\over5}g_1^2{\rm Tr}\{Ym^2\} \nonumber \\
   &&\mbox{}- {4\over15}g_1^2M_1^2
   - {16\over3}g_3^2M_3^2 \;] \ , \label{appeq5} \\
   {dm_{u_i}^2 \over dt} &=& {1\over8\pi^2}[\; \sum_{j}
   2\vert Y_u^{ij}\vert^2 ( m_{\Phi_u}^2 + m_{u_i}^2 + m_{Q_j}^2 +
   \vert A_u^{ij}\vert^2 )
   - {2\over5}g_1^2{\rm Tr}\{Ym^2\} \nonumber \\
   &&\mbox{}- {16\over15}g_1^2M_1^2
   - {16\over3}g_3^2M_3^2 \;] \ , \label{appeq6} \\
   {dm_{Q_i}^2 \over dt} &=& {1\over8\pi^2}[\; \sum_{i,j} \Bigl(
   \vert Y_u^{ji}\vert^2 ( m_{\Phi_u}^2 + m_{Q_i}^2 + m_{u_j}^2 +
   \vert A_u^{ji}\vert^2 )
   + \vert Y_d^{ji}\vert^2 ( m_{\Phi_d}^2 + m_{Q_i}^2 + m_{d_j}^2 \nonumber \\
   &&\mbox{} + \vert A_d^{ji}\vert^2 ) \Bigr)
    + {1\over10}g_1^2{\rm Tr}\{Ym^2\} - {1\over15}g_1^2M_1^2
   - 3g_2^2M_2^2 - {16\over3}g_3^2M_3^2 \;] \ , \label{appeq7} \\
   {dB \over dt} &=& {1\over8\pi^2} [\; 3A_u^{ij}\vert Y_u^{ij}\vert^2
   + 3 A_d^{ij}\vert Y_d^{ij}\vert^2 + A_e^{ij}\vert Y_e^{ij}\vert^2
   - {3\over5}g_1^2M_1 - 3g_2^2M_2 \;] \ .
\end{eqnarray}
where, as in Falck, sums are implied over all indices not appearing on the
left hand side and where
\begin{equation}
   {\rm Tr}\{Ym^2\} = \sum_{i=1}^{n_g} ( m_{Q_i}^2 - 2 m_{u_i}^2 + m_{d_i}^2
                 - m_{L_i}^2 + m_{e_i}^2 ) + m_{\Phi_u}^2 - m_{\Phi_d}^2 \ .
\end{equation}
\narrowtext
\noindent Note that in an anomaly free theory, this term is zero if all the
masses are equal at some scale.  That it is zero at such a scale can
be seen by using the fact that there is no gravitational anomaly.  In such a
case, ${\rm Tr}\{Ym^2\}=m^2{\rm Tr}\{Y\}=0$.
To show that it remains true at all
scales, one also needs the cancellation of the other triangle anomalies.
For example, in Eqs.~(\ref{appeq1}-\ref{appeq7}), the $g_1^2M_1^2$ terms
come from one loop mass corrections involving two bino-particle-sparticle
vertices and are therefore proportional to $Y^2$.  Thus one needs to have
the $U(1)$ anomaly cancellation condition ${\rm Tr}\{Y^3\}=0$ in order to show
$d{\rm Tr}\{Ym^2\}/dt=0$.
\narrowtext
The gaugino masses evolve as follows
\begin{equation}
   {d\ln M_l \over dt} = - {1\over8\pi^2} b_l^{} g_l^2 \ .
\end{equation}

\subsection{ The two loop SUSY $\beta$-functions }

The two loop contributions to the gauge couplings are
\begin{equation}
 (b_{lk}) = \left( \begin{array}{ccc}
                                     {38\over15} & {6\over5} & {88\over15} \\
                                     {2\over5}   & 14        & 8           \\
                                     {11\over15} & 3         & {68\over3}
                    \end{array} \right) n_g +
             \left( \begin{array}{ccc}
                                      {9\over25} & {9\over5} & 0   \\
                                      {3\over5}  & -17       & 0   \\
                                      0          & 0         & -54
                     \end{array} \right) \ ,
\end{equation}
and
\begin{equation}
   (C_{lf}) = \left( \begin{array}{ccc}
                                       {26\over5} & {14\over5} & {18\over5} \\
                                       6          & 6          & 2          \\
                                       4          & 4          & 0
                     \end{array} \right) \ , \,
   {\rm with} \, f = u \ ,\, d \ ,\, e \ ,
\end{equation}

\widetext
The two-loop contributions to the Yukawa couplings are given by
\begin{eqnarray}
   {\bf \beta}_u^{(2)} &=& - 4({\bf Y}_u^\dagger{\bf Y}_u^{})^2
        - 2({\bf Y}_d^\dagger{\bf Y}_d^{})^2 - 2{\bf Y}_d^\dagger
             {\bf Y}_d^{}{\bf Y}_u^\dagger{\bf Y}_u^{}
     - 9{\rm Tr}\{{\bf Y}_u^\dagger{\bf Y}_u^{}\}
         {\bf Y}_u^\dagger{\bf Y}_u^{} \nonumber \\
    &&\mbox{}- {\rm Tr}\{3{\bf Y}_d^\dagger{\bf Y}_d^{}
   +{\bf Y}_e^\dagger{\bf Y}_e^{}\}{\bf Y}_d^\dagger{\bf Y}_d^{}
   - 3{\rm Tr}\{3({\bf Y}_u^\dagger{\bf Y}_u^{})^2
    +{\bf Y}_d^\dagger{\bf Y}_d^{}{\bf Y}_u^\dagger{\bf Y}_u^{}\} \nonumber \\
   &&\mbox{}+ ({2\over5}g_1^2 + 6g_2^2){\bf Y}_u^\dagger{\bf Y}_u^{}
           + ({2\over5}g_1^2){\bf Y}_d^\dagger{\bf Y}_d^{}
   + ({4\over5}g_1^2+16g_3^2){\rm Tr}\{{\bf Y}_u^\dagger
             {\bf Y}_u^{}\} \nonumber \\
   &&\mbox{}+ ({26\over15}n_g+{403\over450})g_1^4 + (6n_g-{21\over2})g_2^4
   + ({32\over3}n_g-{304\over9})g_3^4 \nonumber \\
   &&\mbox{}+ g_1^2g_2^2 + {136\over15}g_1^2g_3^2 + 8g_2^2g_3^2 \\
   {\bf \beta}_d^{(2)} &=& - 4({\bf Y}_d^\dagger{\bf Y}_d^{})^2
          - 2({\bf Y}_u^\dagger{\bf Y}_u^{})^2 - 2{\bf Y}_u^\dagger
         {\bf Y}_u^{}{\bf Y}_d^\dagger{\bf Y}_d^{}
   - 3{\rm Tr}\{{\bf Y}_u^\dagger{\bf Y}_u^{}\}
   {\bf Y}_u^\dagger{\bf Y}_u^{} \nonumber \\
   &&\mbox{}- 3{\rm Tr}\{3{\bf Y}_d^\dagger{\bf Y}_d^{}+{\bf Y}_e^\dagger
        {\bf Y}_e^{}\}{\bf Y}_d^\dagger{\bf Y}_d^{}
   - 3{\rm Tr}\{3({\bf Y}_d^\dagger{\bf Y}_d^{})^2+({\bf Y}_e^\dagger
       {\bf Y}_e^{})^2+{\bf Y}_d^\dagger{\bf Y}_d^{}{\bf Y}_u^\dagger
         {\bf Y}_u^{}\} \nonumber \\
   &&\mbox{}+ ({4\over5}g_1^2){\bf Y}_u^\dagger{\bf Y}_u^{}
   + ({4\over5}g_1^2+6g_2^2){\bf Y}_d^\dagger{\bf Y}_d^{}
   + (-{2\over5}g_1^2+16g_3^2){\rm Tr}\{{\bf Y}_d^\dagger
        {\bf Y}_d^{}\}
    + ({6\over5}g_1^2){\rm Tr}\{{\bf Y}_e^\dagger{\bf Y}_e^{}\} \nonumber \\
    &&\mbox{}+ ({14\over15}n_g+{7\over18})g_1^4 + (6n_g-{21\over2})g_2^4
   + ({32\over3}n_g-{304\over9})g_3^4 + g_1^2g_2^2  + {8\over9}
   g_1^2g_3^2 + 8g_2^2g_3^2 \\
   {\bf \beta}_e^{(2)} &=& - 4({\bf Y}_e^\dagger{\bf Y}_e^{})^2
         - 3{\rm Tr}\{3{\bf Y}_d^\dagger{\bf Y}_d^{}
   + {\bf Y}_e^\dagger{\bf Y}_e^{}\}{\bf Y}_e^\dagger{\bf Y}_e^{}
   - 3{\rm Tr}\{3({\bf Y}_d^\dagger{\bf Y}_d^{})^2+({\bf Y}_e^\dagger
  {\bf Y}_e^{})^2 \nonumber \\
    &&\mbox{}+{\bf Y}_d^\dagger{\bf Y}_d^{}{\bf Y}_u^\dagger{\bf Y}_u^{}\}
   + (6g_2^2){\bf Y}_e^\dagger{\bf Y}_e^{}
   + ({6\over5}g_1^2){\rm Tr}\{{\bf Y}_e^\dagger{\bf Y}_e^{}\}
   + (-{2\over5}g_1^2+16g_3^2){\rm Tr}\{{\bf Y}_d^\dagger{\bf Y}_d^{}\}
   \nonumber \\
   &&\mbox{}+ ({18\over5}n_g+{27\over10})g_1^4 + (6n_g-{21\over2})g_2^4
   + {9\over5}g_1^2g_2^2
\end{eqnarray}

The two-loop contributions to the anomalous dimension of the scalars are
\begin{eqnarray}
   \gamma^{(2)}_{\Phi_u} &=&
   {3\over4}{\rm Tr}\{3({\bf Y}_u^\dagger{\bf Y}_u^{})^2+3{\bf Y}_u^\dagger
      {\bf Y}_u^{}{\bf Y}_d^\dagger{\bf Y}_d^{}\}
   - ({19\over10}g_1^2+{9\over2}g_2^2+20g_3^2){\rm Tr}\{{\bf Y}_u^\dagger
        {\bf Y}_u^{}\} \nonumber \\
   &&\mbox{} - ({279\over800}+{1803\over1600}n_g)g_1^4
            - ({207\over32}+{357\over64}n_g)g_2^4
            - ({27\over80}+{9\over80}n_g)g_1^2 g_2^2 \ , \\
   \gamma^{(2)}_{\Phi_d} &=& {3\over4}{\rm Tr}\{3({\bf Y}_d^\dagger
 {\bf Y}_d^{})^2+3{\bf Y}_d^\dagger{\bf Y}_d^{}{\bf Y}_u^\dagger{\bf Y}_u^{}
   +({\bf Y}_e^\dagger{\bf Y}_e^{})^2\}
   - ({2\over5}g_1^2+{9\over2}g_2^2+20g_3^2){\rm Tr}\{{\bf Y}_d^\dagger
       {\bf Y}_d^{}\} \nonumber \\
         &&\mbox{} - ({9\over5}g_1^2+{3\over2}g_2^2){\rm Tr}\{{\bf Y}_e^\dagger
        {\bf Y}_e^{}\}
    - ({279\over800}+{1803\over1600}n_g)g_1^4
            - ({207\over32}+{357\over64}n_g)g_2^4 \nonumber \\
            &&\mbox{}- ({27\over80}+{9\over80}n_g)g_1^2 g_2^2 \ .
\end{eqnarray}
\narrowtext
\section{ THRESHOLDS }
\label{ threshapp }

Ref.~\cite{mv} gives formulas valid to two loops to compute the
$\beta$-functions of gauge, Yukawa, and scalar self-quartic couplings in a
general gauge theory.  These will be useful in obtaining the required form
of these $\beta$-functions for the purpose of including thresholds.

To implement the super particle thresholds in the minimal low energy super
gravity model, the renormalization group $\beta$-function must be calculated
in a form that allows every particle to be counted in the simplest possible
way.  This, of course, implies that we will have to make allowances for
effective theories with half odd integer doublets, as discussed in Section
VIII.  We will, therefore, implement
particle thresholds as steps in the particle content of the model.
In the following, the one loop $\beta$-functions of the gauge couplings and
Yukawa couplings are considered.

\subsection{ Gauge Couplings }

In the general case (but for a single, simple gauge group $G$), at one loop
\begin{equation}
   {dg\over dt} = {1\over(4\pi)^2} b^{(1)} g^3 \ ,
\label{iieq10a}
\end{equation}
where
\begin{equation}
   b^{(1)} = {2\over3} T_2(F) + {1\over3} T_2(S) - {11\over3} C_2(G) \ ,
\label{iieq10b}
\end{equation}
and where $F$, $S$, and $G$ stand for the fermion, scalar, and adjoint
representations, respectively.
Using the fact that gauginos are in the adjoint representation gives in
the $SU(3)$ case
\begin{equation}
   b_3^{(1)} = {2\over3} T_2(F_3) + {1\over3} T_2(S_3)
                         + {2\over3} C_2(G_3) \theta_{\tilde g}
                         - {11\over3}C_2(G_3) \ ,
\label{threshb1}
\end{equation}
where $F_3$, $S_3$ refer to the fermion and scalar representations, and
$G_3=SU(3)$ and where
\begin{equation}
   \theta_{\tilde g} = \cases{ 1 \ , & $\Lambda > M_g$ \ ; \cr
                               0 \ , & $\Lambda < M_g$ \ . \cr }
\label{deltag}
\end{equation}
$M_g$ is the mass of the gluino.
In the $SU(3)$ case, one has
\begin{equation}
   T_2(R_3) = 2({1\over2}) N_Q + ({1\over2}) N_{\overline u}
              + ({1\over2}) N_{\overline d} \ ,
\label{t2}
\end{equation}
where $N_p$ equals the number of families of particle p.  This result
is valid for both fermion ($R=F$) and scalar ($R=S$) representations.
Equations (\ref{threshb1}) and (\ref{t2}) lead to
\begin{equation}
   b_3^{(1)} = {2\over3} ( N_u + N_d )
             + {1\over3} N_{\tilde Q}
             + {1\over6} N_{\tilde{\overline u}}
             + {1\over6} N_{\tilde{\overline d}}
             + 2 \theta_{\tilde g} - 11 \ ,
\label{newb31lp}
\end{equation}
where we have assumed that $N_Q = (N_u + N_d)/2$.
Also the fact that left and right handed quarks of a given flavor have the
same mass allowed us to use $N_{\overline u}=N_u$ and $N_{\overline d}=N_d$.
Note that Eq.~(\ref{newb31lp}) reduces to the right standard model result when
$N_{\tilde Q} = N_{\tilde{\overline u}} = N_{\tilde{\overline d}} =
\theta_{\tilde g} = 0$, and to the right supersymmetric result
($\Lambda>M_{SUSY}$) when
$N_{\tilde Q}=N_{\tilde{\overline u}}=N_{\tilde{\overline d}}=3$ and
$\theta_{\tilde g}=1$.  Similar formulas are calculated for $g_1$ and $g_2$
\begin{mathletters}
\begin{eqnarray}
   b_1^{(1)} &=& {2\over5}( {17\over12}N_u + {5\over12}N_d
   + {5\over4}N_e + {1\over4}N_\nu ) + {1\over30}N_{\tilde Q}
   + {4\over15}N_{\tilde{\overline u}}
   + {1\over15}N_{\tilde{\overline d}} \nonumber \\
 &&+ {1\over10}N_{\tilde L} + {1\over5}N_{\tilde{\overline e}}
   + {1\over5} ( N_{{\tilde\Phi}_u} + N_{{\tilde\Phi}_d} )
   + {1\over10}( N_{\Phi_u} + N_{\Phi_d} ) \ , \label{newb11lp} \\
   b_2^{(1)} &=& -{22\over3} + {1\over2}( N_u + N_d )
   + {1\over6}( N_e + N_\nu )
   + {1\over2} N_{\tilde Q} + {1\over6} N_{\tilde L} \nonumber \\
 &&+ {1\over3}( N_{{\tilde\Phi}_u} + N_{{\tilde\Phi}_d} )
   + {1\over6}( N_{\Phi_u} + N_{\Phi_d} )
   + {4\over3}\theta_{\tilde W} \ .
   \label{newb21lp}
\end{eqnarray}
\end{mathletters}
The two loop contributions may also be calculated in this manner.

\subsection{ Yukawa Couplings }

We find it useful to define new doublet fields
\begin{mathletters}
\begin{eqnarray}
   \Phi_{h} &=& { }\sin\beta {\tilde\Phi}_u + \cos\beta \Phi_d \ ,
   \label{smhiggs} \\
   \Phi_{H} &=& - \cos\beta \Phi_u - \sin\beta {\tilde\Phi}_d \ ,
   \label{2higgs}
\end{eqnarray}
\end{mathletters}
where $\tan\beta=v_u/v_d$ and ${\tilde\Phi}=i\tau_2\Phi^*$.  The VEVs
of these two new fields are $<\Phi_{h}^0>=v$ and $<\Phi_{H}^0>=0$.
We can now rewrite the Yukawa interaction Lagrangean of the MSSM in
terms of these fields.  The $\beta$-function for the Yukawa couplings
can be computed from formulas in Ref.~\cite{mv}.  At one loop, the
renormalization group equation for a general Yukawa coupling is given by
\begin{equation}
   {d{\bf Y}\over dt} = {1\over(4\pi)^2} {\bf Y} {\bf\beta}^{(1)} \ .
\label{yukrge}
\end{equation}
After replacing
\begin{mathletters}
\begin{eqnarray}
   \Phi_u &=& \sin\beta {\tilde\Phi}_{h} - \cos\beta \Phi_{H} \ ,
   \label{phiu} \\
   \Phi_d &=& \cos\beta \Phi_{h} - \sin\beta {\tilde\Phi}_{H} \ ,
   \label{phid}
\end{eqnarray}
\end{mathletters}
in the up sector Yukawa interaction Lagragian, Eq.~(\ref{yukrge}) yields
\begin{equation}
   {d({\rm s}{\bf Y}_u^{})\over dt} = {1\over(4\pi)^2} ({\rm s}{\bf Y}_u^{})
         {\bf\beta}_{({\rm s}{\bf Y}_u^{})} \ ,
\label{upyukrge}
\end{equation}
where we will use the notation ${\rm s}\equiv\sin\beta$ and
${\rm c}\equiv\cos\beta$.
{}From Eq.~(\ref{upyukrge}), we get
\begin{equation}
   {d{\bf Y}_u^{}\over dt} = {1\over(4\pi)^2} ({\bf Y}_u^{}
             {\bf\beta}_{({\rm s}{\bf Y}_u^{})})
                    - {d\ln{\rm s}\over dt}{\bf Y}_u^{} \ .
\label{upyukrge2}
\end{equation}
Using $\sin\beta=v_u/v$ and Eq.~(\ref{vevrge}) to one loop, we can write
\begin{eqnarray}
   {d\ln{\rm s}\over dt} &=& {\rm c}^2
   ( {{\dot v}_u\over v_u} - {{\dot v}_d\over v_d} ) \nonumber \\
                    &=& {1\over(4\pi)^2} {\rm c}^2
   ( \gamma_{\Phi_u} - \gamma_{\Phi_d} ) \ .
\label{srge}
\end{eqnarray}
Putting Eqs.~(\ref{upyukrge}) and (\ref{srge}) together yields
\begin{equation}
   {d{\bf Y}_u^{}\over dt} = {1\over(4\pi)^2} [ {\bf Y}_u^{}
        {\bf\beta}_{({\rm s}{\bf Y}_u^{})} -
   {\rm c}^2 ( \gamma_{\Phi_u} - \gamma_{\Phi_d} ) {\bf Y}_u^{} ] \ .
\label{uyukfinal}
\end{equation}
Similarly for the down and lepton sectors, we get
\begin{mathletters}
\begin{eqnarray}
   {d{\bf Y}_d^{}\over dt} &=& {1\over(4\pi)^2} [ {\bf Y}_d^{}
     {\bf\beta}_{({\rm c}{\bf Y}_d^{})} +
   {\rm s}^2 ( \gamma_{\Phi_u} - \gamma_{\Phi_d} ) {\bf Y}_u^{} ] \ ,
   \label{dyukfinal} \\
   {d{\bf Y}_e^{}\over dt} &=& {1\over(4\pi)^2} [ {\bf Y}_e^{}
      {\bf\beta}_{({\rm c}{\bf Y}_e^{})} +
   {\rm s}^2 ( \gamma_{\Phi_u} - \gamma_{\Phi_d} ) {\bf Y}_u^{} ] \ .
   \label{eyukfinal}
\end{eqnarray}
\end{mathletters}

\widetext
Now we list the forms of the $\beta$-functions appearing in
Eqs.~(\ref{uyukfinal}), (\ref{dyukfinal}), and (\ref{eyukfinal})
\begin{mathletters}
\begin{eqnarray}
   \Bigl[ {\bf Y}_u^{}{\bf\beta}_{({\rm s}{\bf Y}_u^{})} \Bigr]_{ij} &=&
       \Bigl[ \
   {3\over2} ( {\rm s}^2\theta_{h} + {\rm c}^2\theta_{H} ){\bf Y}_u^{}
     {\bf Y}_u^\dagger{\bf Y}_u^{}
   + {1\over2} ( {\rm s}^2\theta_{\tilde h} + {\rm c}^2\theta_{\tilde H} )
             (\ 2\{\sum_{k=1}^{N_{\tilde Q}}{\bf Y}_u^{}{\bf Y}_u^\dagger\}
       {\bf Y}_u^{} \nonumber \\
 && \mbox{}+  {\bf Y}_u^{}\{\sum_{k=1}^{N_{\tilde{\overline u}}}
         {\bf Y}_u^\dagger{\bf Y}_u^{}\} \ ) +
   {1\over2} ( {\rm c}^2\theta_{h} + {\rm s}^2\theta_{H} -
               4{\rm c}^2( \theta_{h} - \theta_{H} ) ){\bf Y}_u^{}
        {\bf Y}_d^\dagger{\bf Y}_d^{} \nonumber \\
 && \mbox{}+ {1\over2} ( {\rm c}^2\theta_{\tilde h}
         + {\rm s}^2\theta_{\tilde H} )
               {\bf Y}_u^{}\{\sum_{k=1}^{N_{\tilde{\overline d}}}
       {\bf Y}_d^\dagger{\bf Y}_d^{}\} +
   {\bf Y}_u^{} [\ ( {\rm s}^2\theta_{h} + {\rm c}^2\theta_{H} )
        ~{\rm Tr}\{ 3{\bf Y}_u^\dagger{\bf Y}_u^{} \} \nonumber \\
   && \mbox{}+ {\rm c}^2( \theta_{h}
        - \theta_{H} ) ~{\rm Tr}\{ 3{\bf Y}_d^\dagger
     {\bf Y}_d^{} + {\bf Y}_e^\dagger{\bf Y}_e^{} \} \ ]
   - \{\ {3\over5}g_1^2[\ {17\over12} + {3\over4}\theta_{\tilde h} \nonumber \\
   && \mbox{}- ( {1\over36}\theta_{{\tilde Q}_j}
    + {4\over9}\theta_{{\tilde{\overline u}}_i} +
        {1\over4}\theta_{\tilde h} ) \theta_{\tilde B} \ ]
   +   g_2^2[\ {9\over4} + {9\over4}\theta_{\tilde h} - {3\over4}
      ( \theta_{{\tilde Q}_j} + \theta_{\tilde h} ) \theta_{\tilde W} \ ]
   \nonumber \\
   && \mbox{} +   g_3^2[\ 8 - {4\over3} ( \theta_{{\tilde Q}_j}
      + \theta_{{\tilde{\overline u}}_i} )
             \theta_{\tilde g} \ ] \ \} \ \Bigr]_{ij} \ ,
   \label{bfu} \\
   \Bigl[ {\bf Y}_d^{}{\bf\beta}_{({\rm c}{\bf Y}_d^{})} \Bigr]_{ij} &=&
      \Bigl[ \
   {3\over2} ( {\rm c}^2\theta_{h} + {\rm s}^2\theta_{H} )
         {\bf Y}_d^{}{\bf Y}_d^\dagger{\bf Y}_d^{}
   + {1\over2} ( {\rm c}^2\theta_{\tilde h} + {\rm s}^2\theta_{\tilde H} )
             (\ 2\{\sum_{k=1}^{N_{\tilde Q}}{\bf Y}_d^{}{\bf Y}_d^\dagger\}
           {\bf Y}_d^{} \nonumber \\
       && \mbox{}+  {\bf Y}_d^{}\{\sum_{k=1}^{N_{\tilde{\overline d}}}
         {\bf Y}_d^\dagger
          {\bf Y}_d^{}\} \ ) +
   {1\over2} ( {\rm s}^2\theta_{h} + {\rm c}^2\theta_{H} -
               4{\rm s}^2( \theta_{h} - \theta_{H} ) ){\bf Y}_d^{}
            {\bf Y}_u^\dagger{\bf Y}_u^{}  \nonumber \\
   && \mbox{}+ {1\over2} ( {\rm s}^2\theta_{\tilde h}
            + {\rm c}^2\theta_{\tilde H} )
          {\bf Y}_d^{}\{\sum_{k=1}^{N_{\tilde{\overline u}}}{\bf Y}_u^\dagger
         {\bf Y}_u^{}\} +
   {\bf Y}_d^{} [\ ( {\rm s}^2( \theta_{h} - \theta_{H} ) ~{\rm Tr}\{
      3{\bf Y}_u^\dagger{\bf Y}_u^{} \} \nonumber \\
   && \mbox{}+ ( {\rm c}^2\theta_{h} + {\rm s}^2\theta_{H} ) ~{\rm Tr}\{
  3{\bf Y}_d^\dagger{\bf Y}_d^{} + {\bf Y}_e^\dagger{\bf Y}_e^{} \} \ ] -
   \{\ {3\over5}g_1^2[\ {5\over12} + {3\over4}\theta_{\tilde h}   \nonumber \\
     && \mbox{}- ( {1\over36}\theta_{{\tilde Q}_j}
       + {1\over9}\theta_{{\tilde{\overline d}}_i} +
        {1\over4}\theta_{\tilde h} ) \theta_{\tilde B} \ ]
   +   g_2^2[\ {9\over4} + {9\over4}\theta_{\tilde h} - {3\over4}
      ( \theta_{{\tilde Q}_j} + \theta_{\tilde h} ) \theta_{\tilde W} \ ]
    \nonumber \\
   && \mbox{} +  g_3^2[\ 8 - {4\over3} ( \theta_{{\tilde Q}_j}
      + \theta_{{\tilde{\overline d}}_i} )
             \theta_{\tilde g} \ ] \ \} \ \Bigr]_{ij} \ ,
   \label{bfd} \\
   \Bigl[ {\bf Y}_e^{}{\bf\beta}_{({\rm c}{\bf Y}_e^{})} \Bigr]_{ij} &=&
  \Bigl[ \
   {3\over2} ( {\rm c}^2\theta_{h} + {\rm s}^2\theta_{H} ){\bf Y}_e^{}
     {\bf Y}_e^\dagger{\bf Y}_e^{}
   + {1\over2} ( {\rm c}^2\theta_{\tilde h} + {\rm s}^2\theta_{\tilde H} )
             (\ 2\{\sum_{k=1}^{N_{\tilde L}}{\bf Y}_e^{}{\bf Y}_e^\dagger\}
    {\bf Y}_e^{} \nonumber \\
     && \mbox{}+  {\bf Y}_e^{}\{\sum_{k=1}^{N_{\tilde{\overline e}}}
    {\bf Y}_e^\dagger
   {\bf Y}_e^{}\} \ ) +
   {\bf Y}_e^{} [\ ( {\rm s}^2( \theta_{h} - \theta_{H} ) ~{\rm Tr}\{
   3{\bf Y}_u^\dagger{\bf Y}_u^{} \} \nonumber \\
   && \mbox{}+ ( {\rm c}^2\theta_{h} + {\rm s}^2\theta_{H} ) ~{\rm Tr}\{
  3{\bf Y}_d^\dagger{\bf Y}_d^{} + {\bf Y}_e^\dagger{\bf Y}_e^{} \} \ ] -
   \{\ {3\over5}g_1^2[\ {15\over4} + {3\over4}\theta_{\tilde h} \nonumber \\
     && \mbox{}- ( {1\over4}\theta_{{\tilde L}_j}
       + \theta_{{\tilde{\overline e}}_i} +
        {1\over4}\theta_{\tilde h} ) \theta_{\tilde B} \ ]
   +    g_2^2[\ {9\over4} + {9\over4}\theta_{\tilde h} - {3\over4}
      ( \theta_{{\tilde L}_j} + \theta_{\tilde h} ) \theta_{\tilde W} \ ] \ \}
      \ \Bigr]_{ij} \ ,
   \label{bfe}
\end{eqnarray}
\end{mathletters}
\narrowtext
\noindent where the various $\theta$s equal zero below the mass
threshold of the respective particle and equal one above it.  Note
that Eqs.~(\ref{bfu}), (\ref{bfd}), and (\ref{bfe}) reduce to
Eqs.~(\ref{1lpubf}), (\ref{1lpdbf}), and (\ref{1lpebf}), respectively,
in the supersymmetric limit in which all $\theta$s are equal to one.
When $\theta_{h}=1$ and all other $\theta$s in the above equations are
equal to zero, we recover the standard model result after identifying
the standard model Yukawas as ${\bf Y}_u^{\scriptscriptstyle h}={\rm
s}{\bf Y}_u^{}$, ${\bf Y}_d^{\scriptscriptstyle h}={\rm c}{\bf
Y}_d^{}$, and ${\bf Y}_e^{\scriptscriptstyle h}={\rm c}{\bf Y}_e^{}$.

\end{document}